\documentclass[prb,aps,english,twocolumn,notitlepage,superscriptaddress]{revtex4-2}

\usepackage{graphicx}% Include figure files
\usepackage{dcolumn}% Align table columns on decimal point
\usepackage{rotating}
\usepackage{lipsum}
\usepackage{xcolor}
\usepackage{bm}% bold math
\usepackage[T1]{fontenc}  
\usepackage{multirow,amsmath,amssymb}
 \usepackage{babel} 
 \usepackage{txfonts}
 \usepackage{epsfig,epsf,psfrag} 
\usepackage{graphicx} 
\usepackage{pslatex}
\usepackage{epic,eepic} 
\usepackage{color,pstcol}
\usepackage{pstricks} 
\usepackage{fancyhdr} 
\usepackage{tabularx}
\usepackage{physics}
\usepackage{url}
\usepackage{listings}
\usepackage{color} 
\usepackage{caption}
\usepackage{ulem}
\usepackage{balance}
\usepackage{subcaption}
\usepackage{float}
\usepackage[utf8]{inputenc}
\usepackage{comment}

\usepackage{titlesec}
\titleclass{\subsubsubsection}{straight}[\subsubsection]
\newcounter{subsubsubsection}[subsubsection]
\renewcommand\thesubsubsubsection{\thesubsubsection.\arabic{subsubsubsection}}

\titleformat{\subsubsubsection}
  [block] % ← Use 'block' for center alignment
  {\centering\normalfont\normalsize\bfseries} % ← centering added here
  {\thesubsubsubsection} % number format
  {1em} % spacing between number and title
  {}

\titlespacing*{\subsubsubsection}
  {0pt}{3.25ex plus 1ex minus .2ex}{1.5ex plus .2ex}

\usepackage{hyperref}
\hypersetup{
 colorlinks=true,
 linkcolor=blue,
 anchorcolor = blue,
 citecolor = blue,
 filecolor = blue,
 urlcolor = blue
}
\usepackage{dsfont}

\graphicspath{{Images/}}
\def \be {\begin{equation}} 
\def \ee {\end{equation}}

\begin{document}

\preprint{APS/123-QED}
\title{Effect of electron-hole asymmetry on $\Delta_T$ noise in metal/quantum
    point contact/metal and metal/quantum point contact/superconductor
    junctions}
\author{Sachiraj Mishra}
\email{sachiraj29mishra@gmail.com}
\author{Colin Benjamin}
\email{colin.nano@gmail.com}
\affiliation{School of Physical Sciences, National Institute of Science Education and Research, HBNI, Jatni-752050, India}
\affiliation{Homi Bhabha National Institute, Training School Complex, AnushaktiNagar, Mumbai, 400094, India }

\begin{abstract}
{This work examines charge $\Delta_T$ noise in two-terminal hybrid nanostructures featuring a quantum point contact (QPC), realized either between two normal metallic leads (NQN) or between a normal metal and a superconducting lead (NQS). The {energy-dependent transmission of a} QPC breaks electron--hole (e-h) symmetry, leading to a finite {thermovoltage} {under an applied temperature and voltage bias}. In contrast, in earlier studies on hybrid junctions incorporating insulating barriers, as electron–hole symmetry is preserved, have vanishing {thermovoltage}, and consequently, charge $\Delta_T$ noise is calculated at zero {thermovoltage}. In our setup, the broken e-h symmetry allows for a finite {thermovoltage}, at which we compute the corresponding charge $\Delta_T$ noise. Unlike earlier studies restricted by electron–hole symmetry and vanishing {thermovoltage}, our work establishes a self-consistent thermoelectric noise framework in mesoscopic hybrid junctions, revealing how Andreev reflection fundamentally reshapes charge 
$\Delta_T$ noise once electron–hole symmetry is broken. This broad access to charge fluctuation signatures provides a more comprehensive understanding of non-equilibrium transport in linear response. To the best of our knowledge, this is the first work that systematically explores the interplay between Andreev reflection and electron–hole symmetry breaking in the context of quantum noise and $\Delta_T$ noise in mesoscopic hybrid structures.
}

\end{abstract}

\maketitle

\section{Introduction}

{The study of $\Delta_T$ noise has recently attracted considerable attention from both theoretical and experimental perspectives \cite{PhysRevLett.127.136801, PhysRevLett.125.086801, PhysRevB.105.195423, PhysRevB.107.075409, PhysRevB.107.245301, PhysRevLett.125.106801, PhysRevB.107.155405, melcer2022absent, popoff2022scattering, lumbroso2018electronic, shein2022electronic, sivre2019electronic, mishra2025delta_t, PhysRevResearch.7.023321, k95y-7zrb, mishra2025negativespindeltatnoise, shein2024delta, PhysRevB.108.245427, mishra2026delta, mishra2026temperature, Mishra_2026}. Experimental realizations have demonstrated the presence of $\Delta_T$
noise across a broad range of mesoscopic platforms, including single-molecule transport setups \cite{lumbroso2018electronic}, engineered quantum electronic circuits \cite{shein2022electronic}, and metallic tunnel structures subjected to thermal gradients \cite{sivre2019electronic}. $\Delta_T$ noise refers to the quantum shot noise-like component that remains finite when a temperature difference is applied across the system while the net charge current vanishes~\cite{lumbroso2018electronic, PhysRevLett.127.136801, PhysRevB.107.075409}. As such, it serves as a powerful probe of nonequilibrium quantum transport processes driven purely by thermal bias. }

{We investigate charge \( \Delta_T \) noise in normal metal–quantum point contact–normal metal (NQN) and normal metal–quantum point contact–superconductor (NQS) junctions, and present a comparative analysis with the corresponding charge quantum shot noise under finite charge current. In our earlier work~\cite{mishra2025delta_t}, we studied charge \( \Delta_T \) noise, defined at vanishing charge current, in electron–hole (e–h) symmetric systems such as normal metal–insulator–normal metal (NIN) and normal metal–insulator–superconductor (NIS) junctions. In these symmetric setups, the {thermovoltage} vanishes when charge current vanishes, so the \( \Delta_T \) noise is evaluated strictly at this condition.
{In contrast, the NQN and NQS junctions explicitly break e–h symmetry, giving rise to a finite Seebeck coefficient, which characterizes the intrinsic thermoelectric response of the junction and does not depend on the application of a temperature bias. A finite thermovoltage, however, develops only when a temperature bias is applied across the junction under the condition of zero average charge current.
} This asymmetry enables us to investigate how a finite {thermovoltage} influences the charge \( \Delta_T \) noise.  The breaking of e–h symmetry in NQN and NQS junctions thus provides a unique platform to explore charge \( \Delta_T \) noise, offering new insights into nonequilibrium transport phenomena in mesoscopic hybrid systems, insights that are inaccessible in e-h symmetric NIN and NIS junctions.
}

 {This manuscript constitutes the first theoretical investigation of 
$\Delta_T$ noise in a superconducting hybrid junction that explicitly incorporates electron–hole asymmetry. Previous studies of 
$\Delta_T$ noise in superconducting hybrid structures have been restricted to electron–hole symmetric setups, whereas electron–hole asymmetry has so far been considered only in purely normal-metallic contacts, see Refs. \cite{PhysRevLett.127.136801} and \cite{PhysRevB.107.075409}. Refs.~\cite{PhysRevLett.127.136801} and \cite{PhysRevB.107.075409} study 
$\Delta_T$ noise in purely normal-metal junctions with a resonant-tunneling scatterer to probe electron–hole symmetry. In contrast, the present work considers a quantum point contact. Despite the different scatterers, normal junctions show similar qualitative behavior: the {thermovoltage} remains negative, and the resulting $\Delta_T$ noise is always positive. By introducing a quantum point contact that breaks electron–hole symmetry in a superconducting hybrid junction, our work establishes a new and previously unexplored regime for temperature-driven noise. This advance places the present study in a broader context and highlights its distinct contribution to the understanding of nonequilibrium noise phenomena in mesoscopic superconducting systems. In our setup, the energy-dependent transmission of the QPC explicitly breaks
electron-hole symmetry, generating a finite {thermovoltage} at zero average current. This
feature enables us to evaluate $\Delta_T$ noise at finite {thermovoltage}, an
experimentally relevant regime that has remained largely inaccessible in earlier studies
of superconducting hybrid junctions that rely on electron-hole symmetry. This represents
a qualitative advance in the theoretical understanding of noise in superconducting
junctions at the mesoscopic limit.}

In this work, we uncover several key distinctions in the noise characteristics of quantum point contact (QPC)-based NQN and NQS junctions, in contrast to their insulating counterparts, namely, NIN and NIS junctions. When analyzing the \( \Delta_T \) noise in NQN and NQS systems, we observe pronounced oscillations as a function of Fermi energy, features that are absent in the NIN and NIS cases, as reported in Ref.~\cite{mishra2025delta_t}. A similar oscillatory pattern is also observed in quantum shot noise evaluated at finite charge current.
Notably, the enhancement observed for $\Delta_T$ noise when comparing the NQS junction with its NQN counterpart is bounded by a maximum value of 16, consistent with earlier findings for NIS and NIN junctions in the regime of transparent limit~\cite{mishra2025delta_t}. However, the ratio reduces from 16 as temperature increases. This reduction arises from increased thermal excitations at higher temperatures, which promote quasiparticle transport above the superconducting gap. A similar temperature-induced suppression is also observed in the NIN and NIS setups; see \cite{mishra2025delta_t}.

The remainder of this article is organized as follows: In Sec.~\ref{theory}, we outline the theoretical framework underlying our analysis. We begin with a brief overview of the Landauer–Büttiker formalism for mesoscopic systems and then describe quantum transport in both NQN and NQS junctions. This section also presents the general theory of quantum charge noise, followed by the formulation for calculating {thermovoltage}s and charge \( \Delta_T \) noise in the NQN and NQS setups at finite {thermovoltage}. Section~\ref{results} presents our main results on charge \( \Delta_T \) noise, along with {quantum noise} evaluated under finite current induced by a voltage bias. In Sec.~\ref{analysis}, we carry out a detailed comparison of the charge $\Delta_T$ noise as well as {quantum noise} both in the NQN and NQS junctions, including the evaluation of their respective ratios. Finally, Sec.~\ref{conclusion} summarizes the key findings of our study and discusses prospects for experimental realization. Detailed derivations of scattering amplitudes in the NQS junction, expressions for charge current, {thermovoltage}, and quantum and \( \Delta_T \) noise formulations are provided in Appendices~\ref{App_ch_Qn1}–\ref{App_ch_Qn}. The MATHEMATICA code used to compute the conductance, {thermovoltage}, quantum shot noise, and 
$\Delta_T$ noise is provided in Ref.~\cite{github}.

\section{Theory}
\label{theory}
{The Landaeur-Buttiker formalism is a powerful tool for studying transport in systems exhibiting ballistic conduction. According to this approach, charge current is directly proportional to the net transmission probability of electron scattering between terminals. This formalism has proven invaluable for analyzing various junctions, including QPCs \cite{BENENTI20171, PhysRevLett.60.848, wharam1988one, van1992thermo, PhysRevLett.68.3765, noise} in normal metals, NIN and NIS junctions. In the following sections, we delve into the details of these configurations to understand the transport properties and underlying physical mechanisms.}

\subsection{Landauer-Büttiker Formalism}

In the next two subsections, we specifically discuss the charge current formula in two terminal junctions, first normal metal-QPC-normal metal junction (NQN) and then normal metal-QPC-superconductor (NQS) junction.

\subsubsection{NQN Junction}

Using the Landauer–Büttiker approach, the average charge current at terminal 
$\alpha$ ($\langle I_{\alpha} \rangle$) in a multiterminal normal-state conductor is written as~\cite{BENENTI20171, datta_1995}:
\begin{equation}
\begin{aligned}
\langle I_{\alpha} \rangle &= \frac{2q}{h} \sum_{\beta} \int_{-\infty}^{\infty} dE' \, (\delta_{\alpha \beta} - \mathcal{T}_{\alpha \beta}) \left[ f_{\alpha}(E') - f_{\beta}(E') \right], \\
% \langle J_{\alpha} \rangle &= \frac{2}{h} \sum_{\beta} \int_{-\infty}^{\infty} dE' \, (E' - \mu_{\alpha}) \, (\delta_{\alpha \beta} - \mathcal{T}_{\alpha \beta}) \left[ f_{\alpha}(E') - f_{\beta}(E') \right],
\end{aligned}
\label{eqn:IJ}
\end{equation}

where $\mathcal{T}_{\alpha\beta}(E')$ is the energy-dependent transmission probability from terminal $\beta$ to $\alpha$, where $E'$ is the total energy and $f_{\alpha}(E')$ is the Fermi-Dirac distribution in terminal $\alpha$, given by
\begin{equation}
f_{\alpha} = \frac{1}{1 + e^{(E' - \mu_{\alpha}) / k_B T_{\alpha}}}.
\label{eqn:F}
\end{equation}

Here, $\mu_{\alpha} = E_F + q V_{\alpha}$ is the chemical potential in terminal $\alpha$. $E_F$ denotes the Fermi level, $V_{\alpha}$ represents the applied voltage bias in terminal $\alpha$, and $T_{\alpha} = T + \Delta T_{\alpha}$ is the temperature of terminal $\alpha$, with $T$ taken as the reference temperature of the system and $\Delta T_{\alpha}$ describing the imposed thermal offset at terminal $\alpha$. The quantity $q$ specifies the quasiparticle charge, taking the value $q = e$ for electron-like excitations and $q = -e$ for hole-like excitations.

In the linear response regime, i.e., when $qV_{\alpha} \ll {k_BT}$ and $\Delta T_{\alpha} \ll T$, the charge current can be expressed as~\cite{datta_1995, BENENTI20171}:
\begin{equation}
\langle I_{\alpha} \rangle = \sum_{\beta} G_{\alpha \beta} V_{\beta} + \Pi_{\alpha \beta} \Delta T_{\beta}.
\label{eqn:IG_mat}
\end{equation}

The transport coefficients in this expression are defined as:
\begin{equation}
\begin{split}
& \text{Charge conductance:}\,\,\,   G_{\alpha \beta}\\& \quad \quad \quad \quad \quad \quad \quad \,\,\, = \frac{2q^2}{h} \int_{-\infty}^{\infty} dE \,\,\, (\delta_{\alpha \beta} - \mathcal{T}_{\alpha \beta}) \, \, \, \left(-\partial f/\partial E\right),\\
 & \text{Seebeck coefficient: } S_{\alpha \beta}= \,\,\, \frac{\Pi_{\alpha \beta}}{G_{\alpha \beta}}\,\,\,\text{with,}\\
 &\quad \quad \quad \quad \quad \, \,\,\, \,\,\,\,\,\, \Pi_{\alpha \beta} = \frac{2q}{h T} \int_{-\infty}^{\infty} dE \,\,\, (\delta_{\alpha \beta} - \mathcal{T}_{\alpha \beta}) \,\,\, (E) \left(-\partial f/\partial E\right),\\
    \end{split}
\label{eqn:IG}
\end{equation}

{where, $E$ is the excitation energy, i.e., $E = E' - E_F$ , $f = \left(1 + e^{E/k_B T}\right)^{-1}$ is the equilibrium Fermi-Dirac distribution and $h$ being the Planck's constant.}

\begin{figure}[H]
\centering

\begin{subfigure}{0.80\linewidth}
\centering
\includegraphics[width=\linewidth,height=0.9\textheight,keepaspectratio]{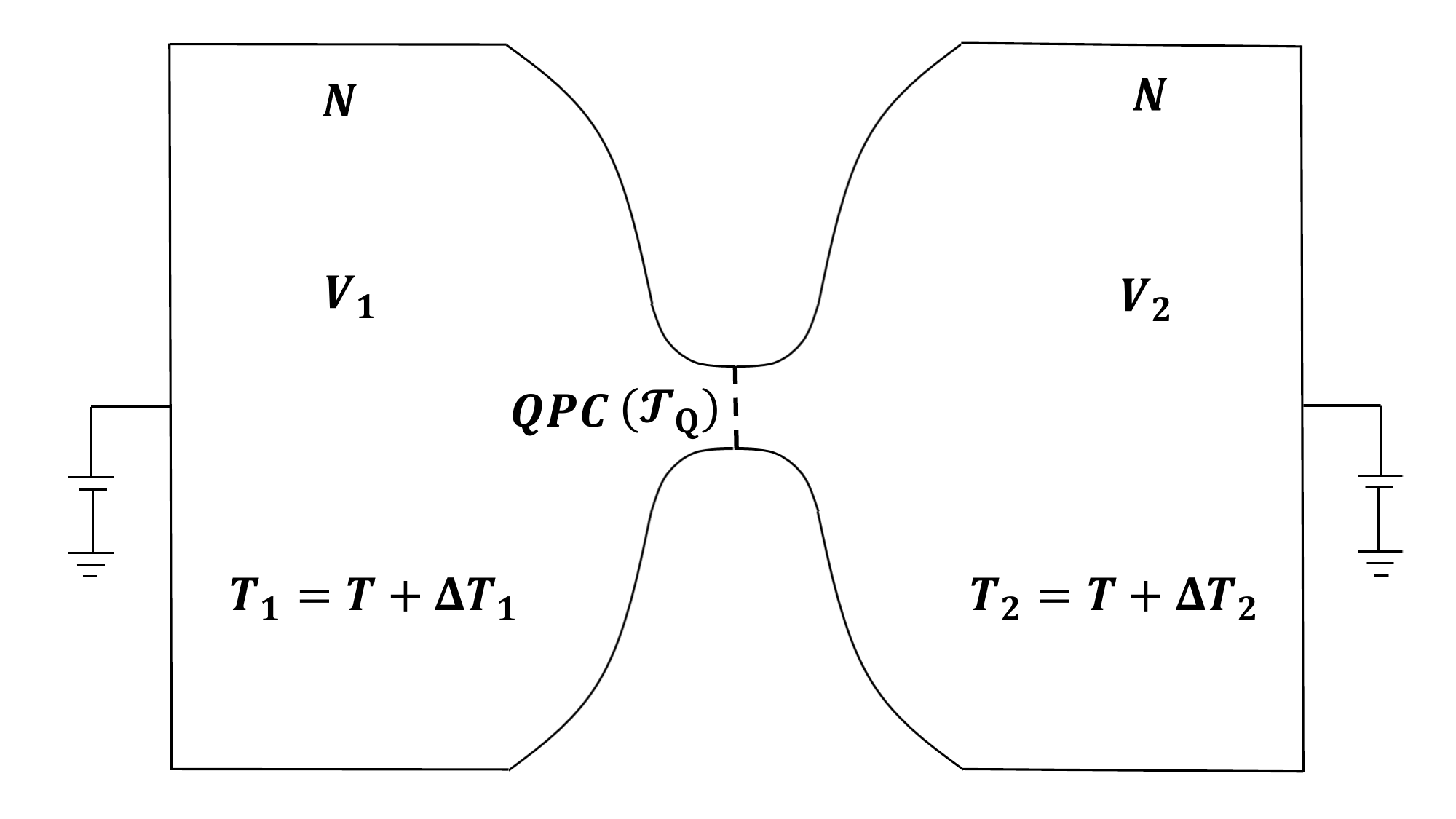}
\caption{}
\end{subfigure}
\\
\begin{subfigure}{0.80\linewidth}
\centering
\includegraphics[width=\linewidth,height=0.9\textheight,keepaspectratio]{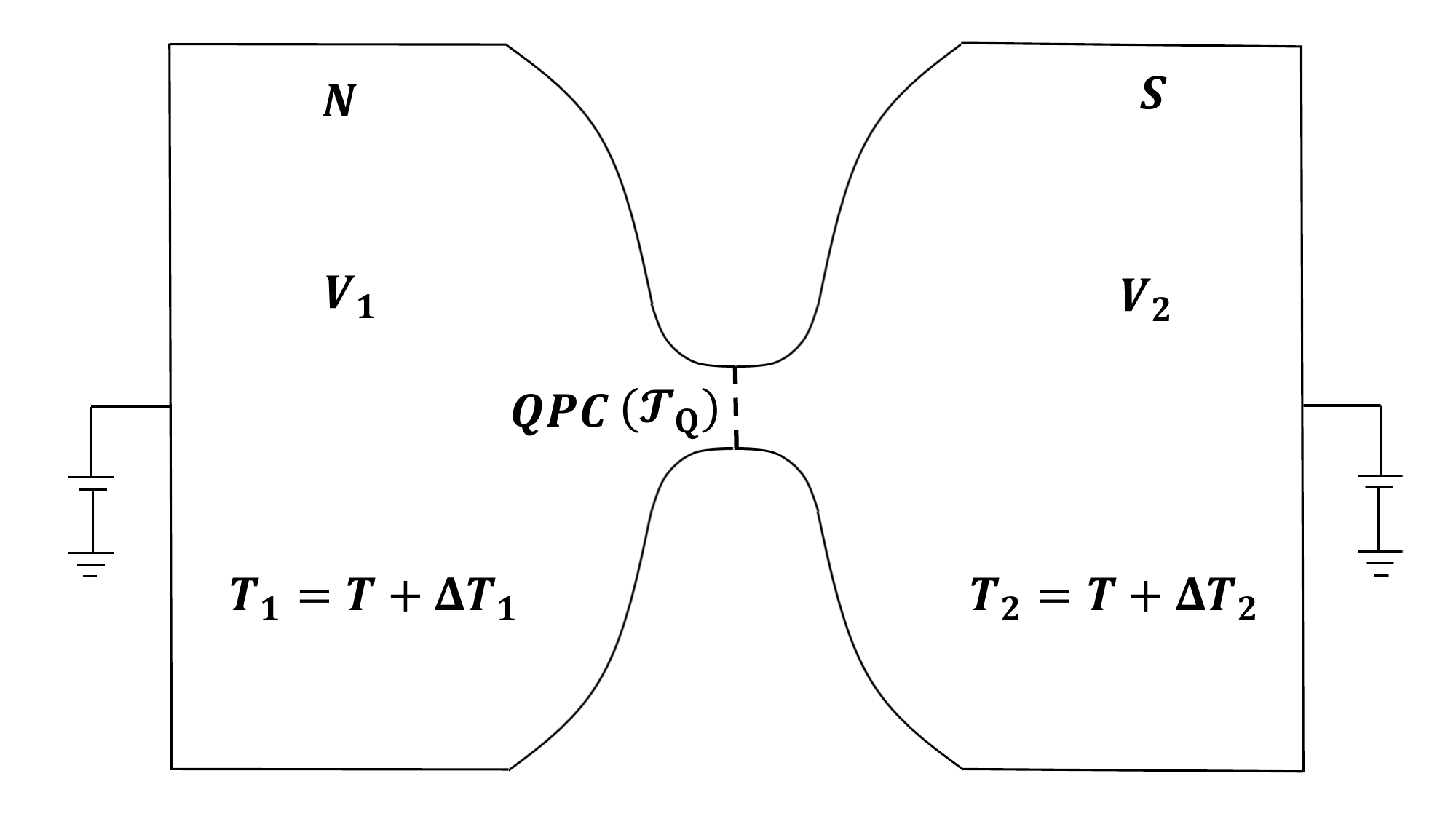}
\caption{}
\end{subfigure}
\caption{Schematic diagram of the (a) NQN junction and (b) NQS junction. The black dashed line in the middle represents the QPC constriction. We consider $V_1 = \Delta V$, $V_2 = 0$, $\Delta T_1 = - \Delta T_2 = \Delta T/2$.}
\label{fig1}
\end{figure}

{In an NQN junction, see Fig. \ref{fig1}(a), the transmission probability through the QPC explicitly breaks electron-hole symmetry as it is asymmetric with respect to energy; in contrast, for a NIN junction, electron-hole symmetry ensures that transmission probability is an even function of energy.} As a result, in a NIN {junction}, the off-diagonal Onsager coefficients such as $\Pi_{\alpha \beta}$ vanish, leading to zero charge current under zero applied voltage bias~\cite{mishra2025delta_t}. However, in a NQN configuration, $\Pi_{\alpha \beta}$ is finite, allowing for finite {thermovoltage} in the absence of net charge current.

\begingroup

The quantum point contact (QPC) has been extensively studied since the pioneering experimental demonstrations of conductance quantization in two-dimensional electron gases in~\cite{PhysRevLett.60.848} and ~\cite{wharam1988one}. Subsequently, both theoretical and experimental thermoelectric properties of QPCs were investigated in Refs.~\cite{van1992thermo,PhysRevLett.68.3765}. In the present study, we employ the widely used saddle-point method to model the QPC, see Ref.~\cite{PhysRevB.41.7906}, in which the confinement potential is given by:
\begin{equation}
U(x, y) = \frac{1}{2}m^* \omega_y^2 y^2 - \frac{1}{2}m^* \omega_x^2 x^2 + V_0,
\end{equation}
where $m^*$ is the electron's effective mass, $V_0$ is the potential offset, $\omega_x$ is the frequency that is responsible for the energy values of each level, whereas $\omega_y$ is the frequency that is responsible for the width of QPC levels. The energy levels for a QPC are quantized as $E_n = V_0 + (n + 1/2)\hbar \omega_y$. The corresponding transmission probability for the $n$th mode as derived in Ref. \cite{PhysRevB.41.7906} is given as,
\begin{equation}
\mathcal{T}_{Q} (E) = \left(1 + e^{-2\pi(E + E_F - E_n)/\hbar \omega_x}\right)^{-1}.
\label{eq61}
\end{equation}

\endgroup

To compute the charge current in the NQN junction as shown in Fig. \ref{fig1}, we consider a voltage bias applied in the left normal metal as $V_1 = \Delta V$ and in the right normal metal as $V_2 = 0$. The left normal-metal lead is maintained at a temperature $T_1 = T + \Delta T_1$, while the right normal-metal lead is held at $T_2 = T + \Delta T_2$, where $\Delta T_1 = -\Delta T_2 = \Delta T/2$, see Fig. \ref{fig1}. An electron incident from the left normal-metal lead is reflected back with probability $1 - \mathcal{T}_{Q}$, while its transmission across the QPC into the right lead occurs with probability $\mathcal{T}_{Q}(E)$. Thus, according to Eq. (\ref{eqn:IG_mat}), the charge current in terminal 1 is given by
\begin{equation}
\langle I_1^{NQN} \rangle = G_{NQN} \Delta V + \Pi_{NQN} \Delta T,
\label{eqn:I_NQN}
\end{equation}
where, for multichannel ($n > 0$),

\begin{equation}
\begin{split}
&  G_{NQN}=\frac{2e^2}{h} \int_{-\infty}^{\infty} dE \,\,\, \sum_{n}\mathcal{T}_{Q}(E) \, \, \, \left(-\partial f/\partial E\right),\\
 & \Pi_{NQN} = \frac{2e}{h T} \int_{-\infty}^{\infty} dE \,\,\, \sum_{n} \mathcal{T}_{Q}(E) E\left(-\partial f/\partial E\right).
    \end{split}
\label{eqn10}
\end{equation}

Here, the charge current $\langle I_{1}^{NQN} \rangle$ vanishes at a finite {thermovoltage}, given by
$V_{th}^{NQN} = \frac{-\Pi_{NQN} }{G_{NQN}}\Delta T = -S_{NQN} \Delta T$. $S_{NQN}$ is the Seebeck coefficient in the NQN junction.
For multichannel transport ($n > 0$), the zero-temperature charge conductance becomes $
G_{NQN} = \frac{2e^2}{h} \sum_n \mathcal{T}_{Q}$.
We now turn to the NQS junction and discuss the general formula for charge current as well as {thermovoltage}.

\subsubsection{NQS Junction}

Similarly, in a multiterminal normal metal-superconductor junction, the charge current out of a normal metallic terminal is given as \cite{BENENTI20171},

\begin{equation}
    \begin{split}
        \langle I_{\alpha} \rangle &= \frac{2 e }{h} \sum_{\beta} \sum_{\rho, \xi \in \{e, h \}} \int_0^{\infty} dE \,\,\, sgn(\rho) \left[\delta_{\alpha \beta} \delta_{\rho \xi} - \mathcal{T}_{\alpha \beta}^{\rho \xi}\right] f_{\beta}^{\xi}(E).
        % \langle J_{\alpha} \rangle &= \frac{2}{h} \sum_{\beta} \sum_{\rho, \xi \in \{e, h \}} \int_0^{\infty} dE \,\,\, (E - sgn(\rho) V_{\alpha}) \left[\delta_{\alpha \beta} \delta_{\rho \xi} - \mathcal{T}_{\alpha \beta}^{\rho \xi}\right] f_{\beta}^{\xi}(E).
    \end{split}
\end{equation}

where $f_{\beta}^{\xi} (E) = \frac{1}{1 + e^{\frac{E - sgn(\xi) V_{\beta}}{k_B T_{\beta}}}}$, where $sgn(\xi)$ = +1 for electron and -1 for hole. Within the linear response framework, the charge current can be written in an analogous manner to Eq.~(\ref{eqn:IG_mat}), where the Onsager coefficients are ~\cite{BENENTI20171},

\begin{equation}
\begin{split}
& \text{Charge conductance:}  \\
&  G_{\alpha \beta} = \frac{2e^2}{h} \sum_{\rho, \xi \in \{e, h \}}  \int_{0}^{\infty} dE \,\,\, sgn(\rho) sgn(\xi) (\delta_{\alpha \beta} \delta_{\rho \xi} - \mathcal{T}_{\alpha \beta}^{\rho \xi}) \left(-\frac{\partial f}{\partial E}\right),\\
 & \text{Seebeck coefficient:} \,\,\, S_{\alpha \beta} = \,\,\, \frac{\Pi_{\alpha \beta}}{G_{\alpha \beta}}\,\,\,\text{with,}\\
 &  \Pi_{\alpha \beta} = \frac{2e}{h T} \sum_{\rho, \xi \in \{e, h \}} \int_{0}^{\infty} dE \,\,\, sgn(\rho) (\delta_{\alpha \beta} \delta_{\rho \xi} - \mathcal{T}_{\alpha \beta}^{\rho \xi}) \,\,\, (E) \left(-\frac{\partial f}{\partial E}\right).\\
    \end{split}
\label{eqn:IG1}
\end{equation}

We now consider a normal metal–quantum point contact–superconductor (NQS) junction, see Fig. \ref{fig1}(b). The relevant scattering amplitudes and probabilities are derived in Appendix A, following Refs.~\cite{PhysRevB.46.12841, RevModPhys.69.731, PhysRevB.99.134514}.

The scattering processes in this hybrid junction include Andreev reflection probability: $\mathcal{R}^A(E)  = |s_{11}^{he}|^2$, normal reflection probability: $\mathcal{R}^B (E) = |s_{11}^{ee}|^2$, with the transmission probabilities $\mathcal{T}^C(E)  = |s_{21}^{ee}|^2$ and $\mathcal{T}^D(E)  = |s_{21}^{he}|^2$ associated with electron-like and hole-like excitations, respectively.

\begin{figure}[H]
\centering
\includegraphics[scale=0.30]{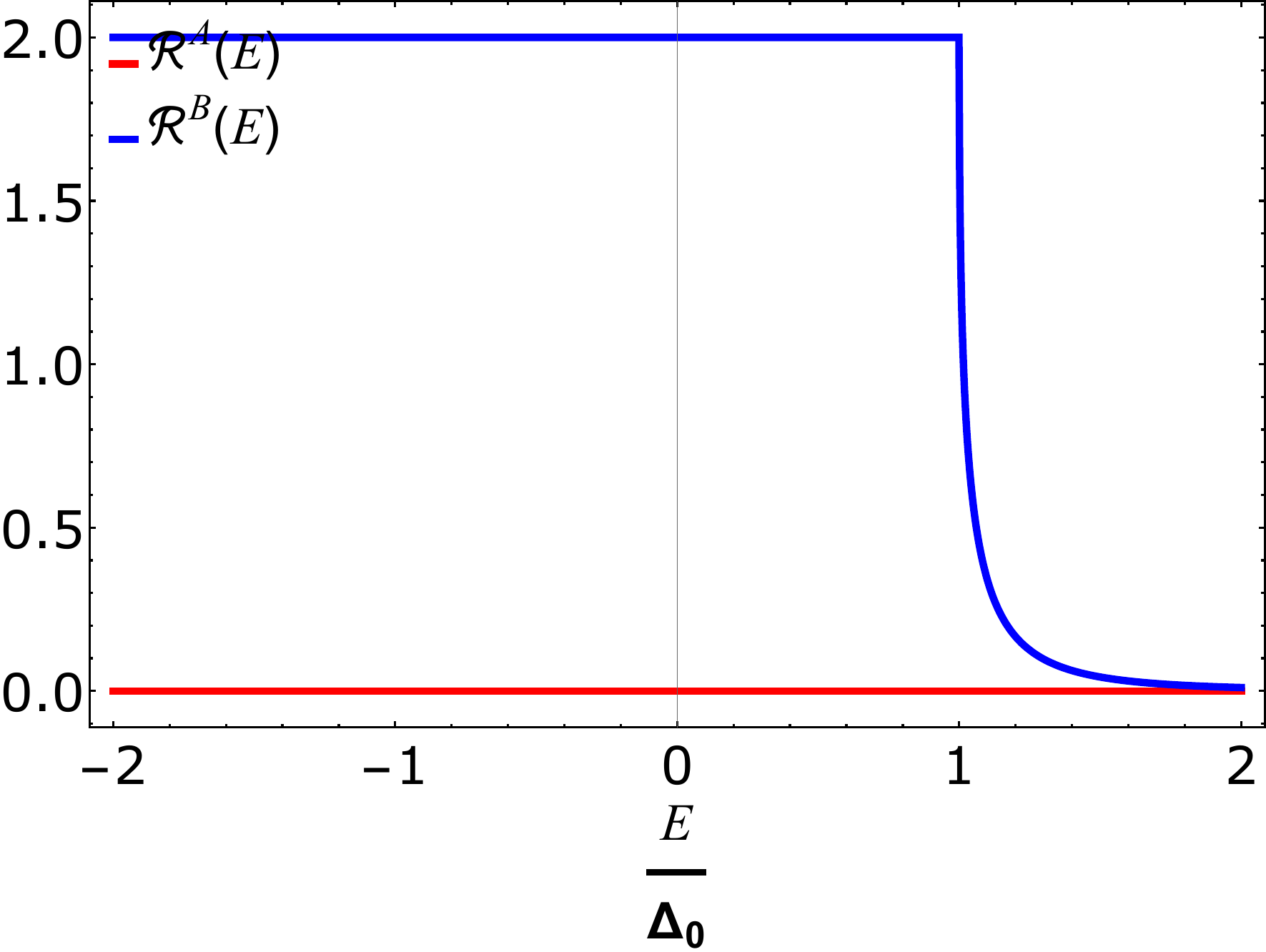}
\caption{$\mathcal{R}^{A}(E)$ and $\mathcal{R}^{B}(E)$ vs. $\frac{E}{\Delta_0}$. Parameters taken are $ \hbar \omega_y = \Delta_0/3$, $\hbar \omega_x =0.1 \hbar \omega_y$, $n = 0,1$, $E_F = V_0 + 1.2 \hbar \omega_x$.}
\label{fig20}
\end{figure}

\begingroup

From Appendix \ref{App_ch_Qn1}, $\mathcal{R}^A(E) = |s_{11}^{he}(E)|^2$ and $\mathcal{R}^B(E) = |s_{11}^{ee}(E)|^2$, where $s_{11}^{he}(E)$ is the Andreev reflection amplitude for electron incident and $s_{11}^{ee}(E)$ is the normal reflection amplitude for the electron incident, $s_{11}^{ee}(E) = \frac{i \left(\sqrt{1-\mathcal{T}_{Q}(E)}-a^2 \sqrt{1-\mathcal{T}_{Q}(-E)}\right)}{D}$ and $s_{11}^{he}(E) = -\frac{a \sqrt{\mathcal{T}_{Q}(E)} \sqrt{\mathcal{T}_{Q}(-E)}}{D}$, where $D = a^2 \sqrt{1 - \mathcal{T}_{Q}(E)} \sqrt{1 - \mathcal{T}_{Q}(-E)} - 1$, where $a=\frac{v}{u}$ and $u=\frac{1}{\sqrt{2}}\left(1+\sqrt{\frac{E^2-\Delta_0^2}{E^2}}\right)^{\frac{1}{2}}, \quad
v=\frac{1}{\sqrt{2}}\left(1-\sqrt{\frac{E^2-\Delta_0^2}{E^2}}\right)^{\frac{1}{2}}$ are the coherence factors in the superconductor, and $\Delta_0=1.76\,k_B T_C \simeq 1.365 meV$, where $T_C = 9K$ denotes the superconducting critical temperature, corresponding to a large-gap $s$-wave superconductor such as Nb\cite{PhysRevB.24.1200}. In this work, all numerical calculations are performed at the temperatures $T=0.01$, $0.1$, $1$, $3$, and $5K$ for $T_C = 9K$ for Niobium \cite{tinkham2004introduction}. At these temperatures, the superconducting gap remains nearly constant and is well approximated by, $\Delta(0)=1.76k_B T_C$ \cite{tinkham2004introduction}.

The QPC transmission probability from Eq. (\ref{eq61}) is $\mathcal{T}_Q(E) =\bigg(1 + e^{-2\pi(E + E_F-V_0 - (n+\frac{1}{2})\hbar \omega_y)/\hbar \omega_x}\bigg)^{-1}$, where $n$ denotes the $n$th energy level. The coherence factors $u$ and $v$ depend only on $E^2$. Consequently, $
u(E)=u(-E),
v(E)=v(-E),$
which implies $
a(E)=\frac{v(E)}{u(E)}=a(-E).$ Therefore, one can write $a(E) = a(-E) = a$.

\begin{figure}[H]
\centering
\includegraphics[scale=0.30]{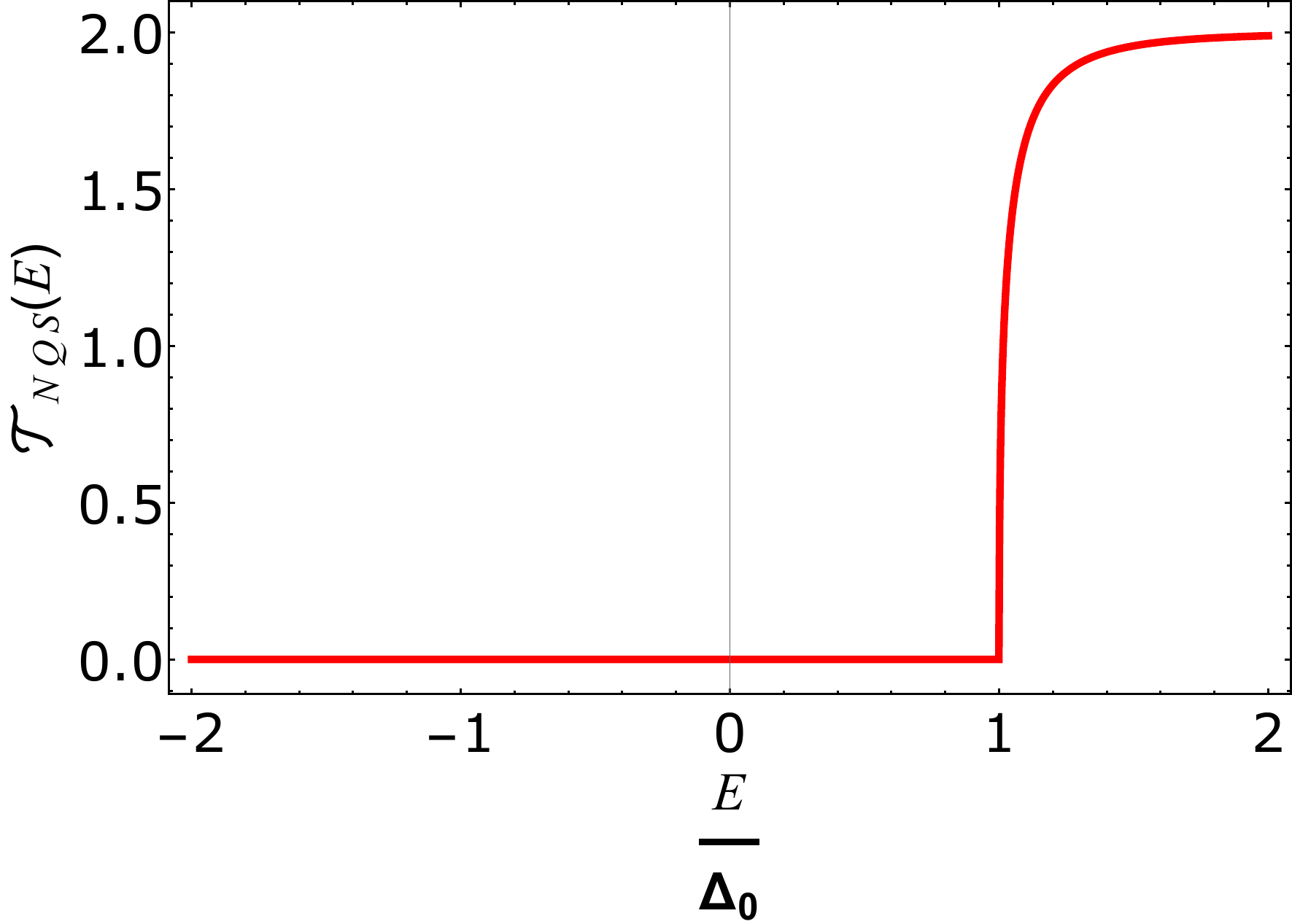}
\caption{$\mathcal{T}_{NQS}(E)$ vs. $\frac{E}{\Delta_0}$. Parameters taken are $ \hbar \omega_y = \Delta_0/3$, $\hbar \omega_x =0.1 \hbar \omega_y$, $n = 0,1$, $E_F = V_0 + 1.2 \hbar \omega_x$.}
\label{fig21}
\end{figure}

    The analytical expressions of $\mathcal{R}^A(E)$ and $\mathcal{R}^B(E)$ in both $E<\Delta_0$ and $E\geq \Delta_0$ regimes are given as, 
    \begin{equation}
    \begin{split}
    \mathcal{R}^{A}(E)&=
\frac{|a|^2\,\mathcal{T}_Q(E)\,\mathcal{T}_Q(-E)}
{\left|1-a^2\sqrt{1-\mathcal{T}_Q(E)}
\sqrt{1-\mathcal{T}_Q(-E)}\right|^2},\\ \quad\mathcal{R}^{B}(E) &= \frac{
\left|
\sqrt{1-\mathcal{T}_Q(E)}
-
a^2\sqrt{1-\mathcal{T}_Q(-E)}
\right|^2
}{
\left|
1-a^2\sqrt{1-\mathcal{T}_Q(E)}
\sqrt{1-\mathcal{T}_Q(-E)}
\right|^2
}.
\end{split}
\label{eqtrans}
  \end{equation}  
    
 $\mathcal{R}^A(E)$ as in Eq. (\ref{eqtrans}) is symmetric with $E$. However, we notice that $\mathcal{R}^B(E)$ is not symmetric under the transformation $E \to -E$ in general, but for $|E|<\Delta_0$, $\mathcal{R}^B(E)$ is symmetric with $E$. The net transmission probability across the NQS junction for arbitrary channel $n$ is given as~\cite{PhysRevB.25.4515} $
\mathcal{T}_{NQS} (E) = 1 + \mathcal{R}^A (E) - \mathcal{R}^B (E) $ and for an arbitrary channel $n$, it is given as

\begin{widetext}

\begin{equation}
    \mathcal{T}_{NQS}(E) =  \frac{\mathcal{T}_{Q} (E) - a^4 \mathcal{T}_{Q} (E) (1 - \mathcal{T}_{Q}(-E)) + \mathcal{T}_{Q}(E) \mathcal{T}_{Q}(-E) |a|^2}{\left(1 - \sqrt{1 - \mathcal{T}_{Q}(E)} \sqrt{1 - \mathcal{T}_{Q}(-E)} a^2\right)^2},
\end{equation}
\end{widetext}

When we plot both $\mathcal{R}^A(E)$ and $\mathcal{R}^B(E)$ as a function of $\frac{E}{\Delta_0}$ for 2 energy levels of QPC, i.e., $n=0,1$ and $\hbar \omega_y = \Delta_0/3, \hbar \omega_x = 0.1 \hbar \omega_y$. In this regime, the normal reflection probability $\mathcal{R}^{B}(E)$ exhibits a distinct energy dependence, as shown in Fig. \ref{fig20}. Within the superconducting gap ($|E|<\Delta_0$), $\mathcal{R}^{B}(E)$ is symmetric with respect to $E$ as expected, reflecting the particle-hole symmetry of subgap transport. In contrast, for energies above the superconducting gap ($|E|>\Delta_0$), $\mathcal{R}^{B}(E)$ is strongly asymmetric, see again Fig. \ref{fig20}. On the negative-energy side ($E<-\Delta_0$), $\mathcal{R}^{B}(E)$ remains constant, whereas on the positive-energy side ($E>\Delta_0$), it decreases rapidly. This pronounced asymmetry originates from the imbalance between electron-like and hole-like quasiparticle excitations outside the superconducting gap. In contrast, the Andreev reflection probability, $\mathcal{R}^{A}(E)$, remains zero with respect to $E/\Delta_0$. However, in any other regime, it will be symmetric. Consequently, the total transmission probability, $\mathcal{T}_{\mathrm{NQS}}(E)$, inherits these characteristics: it is symmetric within the superconducting gap, but becomes increasingly asymmetric in the above-gap regime due to the asymmetric behavior of $\mathcal{R}^{B}(E)$, as seen in Fig.~\ref{fig21}.
\endgroup

Now, similar to NQN junction, the average charge current $\langle I_1^{NQS} \rangle$ in the normal metal is expressed as
\begin{equation}
\langle I_1^{NQS} \rangle = G_{NQS} \Delta V + \Pi_{NQS} \Delta T,
\label{eqn:IG_mat_NIS}
\end{equation}

where for multichannel transport ($n>0$),

\begin{equation}
\begin{split}
&  G_{NQS}=\frac{2e^2}{h} \int_{-\infty}^{\infty} dE \,\,\, \sum_{n}\mathcal{T}_{NQS}(E) \, \, \, \left(-\partial f/\partial E\right),\\
 & {\Pi_{NQS} = \frac{2e}{h T} \int_{-\infty}^{\infty} dE \,\,\, \sum_{n}\mathcal{T}_{NQS}(E) E \left(-\partial f/\partial E\right)}.
    \end{split}
\label{eqn11}
\end{equation}

\begin{widetext}

\begin{figure}[H]
\centering
\includegraphics[scale=0.30]{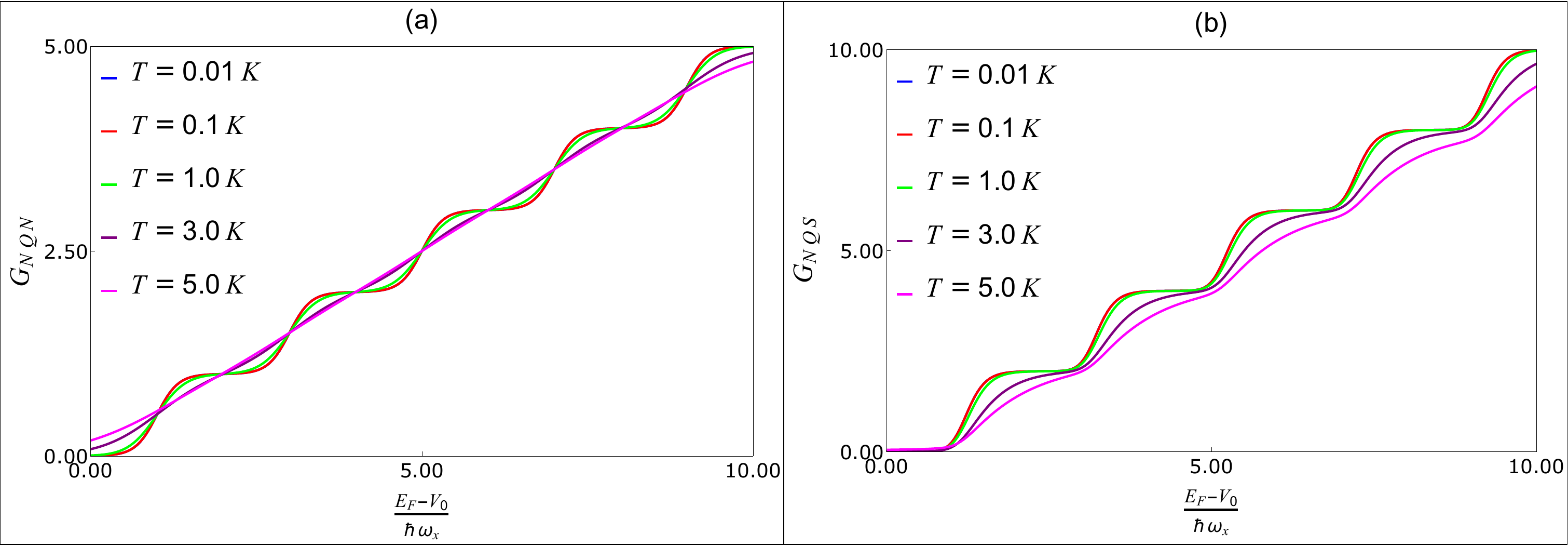}
\caption{Charge conductance in (a) NQN ($G_{NQN}$) and (b) NQS ($G_{NQS}$) (in units of $\frac{2e^2}{h}$) vs. normalized Fermi energy ($\frac{E_F - V_0}{\hbar \omega_x}$) at different temperatures considered above. Parameters taken are $ \hbar \omega_y = 1.365 meV$, $\hbar \omega_x =0.5 \hbar \omega_y$, where $T_C = 9K$. {The curves corresponding to the temperatures $T = 0.01K,0.1K$ nearly overlap and are therefore difficult to distinguish on the scale of the figure}.}
\label{fig3}
\end{figure}

% \begin{figure}[H]
% \centering
% \includegraphics[scale=0.30]{fig4.pdf}
% \caption{Thermal conductance in (a) NQN ($K_{NQN}$) and (b) NQS ($K_{NQS}$) (in units of $\frac{2}{hT}(k_B T)^2$) vs. normalized Fermi energy ($\frac{E_F - V_0}{\hbar \omega_x}$) at different temperatures considered above. The parameters taken are $ \hbar \omega_y = 1.365 meV$, $\hbar \omega_x =0.5 \hbar \omega_y$, where $T_C = 9K$.}
% \label{fig4}
% \end{figure}
\end{widetext}

In the NQN case, the charge current vanishes at a finite {thermovoltage} $V_{NQS}^{th} = \frac{-\Pi_{NQS}}{G_{NQS}}\Delta T = -S_{NQS} \Delta T$. $S_{NQS}$ is the Seebeck coefficient in the NQS junction.
In Fig.~\ref{fig3}, we present the conductance profiles \( G_{NQN} \) and \( G_{NQS} \) as functions of the normalized Fermi energy \( \frac{E_F - V_0}{\hbar \omega_x} \) for the transverse mode index \( n = 4 \), evaluated at various temperatures: \( T = 0.01K, 0.1K, 1.0K,  3.0\,K, \) and \( 5.0\,K \). At $T = 0.01K, 0.1K, 1.0K$, both \( G_{NQN} \) and \( G_{NQS} \) exhibit well-defined quantized conductance plateaus as a function of \( \frac{E_F - V_0}{\hbar \omega_x} \). As the temperature increases to \( T = 3.0\,K \) and further to \( T = 5.0\,K \), thermal broadening leads to a gradual smearing of these steps, and the conductance plateaus begin to diminish, eventually disappearing at higher temperatures. Notably, throughout the entire temperature range, the conductance \( G_{NQS} \) in the NQS (normal-metal–quantum-point-contact–superconductor) junction remains consistently twice that of the \( G_{NQN} \) in the NQN (normal-metal–quantum-point-contact–normal-metal) junction. This doubling is attributed to Andreev reflection at the superconductor interface, where each incident electron is retroreflected as a hole, effectively contributing twice the charge to the conductance compared to the NQN case.

In the next section, we calculate the quantum noise as well as $\Delta_T$ noise in both NQN and NQS junctions.

\subsection{Quantum Noise}

Quantum noise, particularly the auto-correlation of charge current, plays a vital role in understanding fluctuation phenomena in mesoscopic systems. This section is focused on the charge quantum noise in NQN and NQS junctions, with particular emphasis on the zero-frequency limit.

\subsubsection{NQN}

This section presents an analysis of the charge quantum noise in the NQN junction, with the derivation provided in Appendix~\ref{App_ch_Qn}. The total quantum noise is further decomposed into two distinct components: one is quantum {equilibrium noise}, and another is quantum shot noise.

{It is useful to distinguish between quantum equilibrium noise and quantum shot noise. Quantum equilibrium noise refers to the equilibrium current fluctuations (under zero voltage and temperature bias) arising only at finite temperature. In contrast, quantum shot noise originates from the discrete nature of charge carriers under a finite voltage bias when a finite charge current flows through the junction.}

The quantum noise autocorrelation characterizes fluctuations of the charge current flowing through the normal-metal lead of an NQN junction. More broadly, current–current correlations describing charge noise between terminals $i$ and $j$, evaluated at times 
$t$ and $\bar{t}$ are defined as~\cite{PhysRevB.53.16390, noise} $
Q_{ij}(t - \bar{t}) \equiv \langle \Delta I_i(t) \Delta I_j(\bar{t}) + \Delta I_j(\bar{t}) \Delta I_i(t) \rangle$,
where $\Delta I_i(t) = I_i(t) - \langle I_i(t) \rangle$ denotes the current fluctuation at terminal $i$.
Taking the Fourier transform yields the charge noise power spectrum:
$
\delta(\omega + \bar{\omega}) Q_{ij}(\omega) \equiv \frac{1}{2\pi} \langle \Delta I_i(\omega) \Delta I_j(\bar{\omega}) + \Delta I_j(\bar{\omega}) \Delta I_i(\omega) \rangle
$, where $\omega$ and $\bar{\omega}$ are the frequencies associated with energy difference between incoming particles at two different times $t$ and $\bar{t}$ meaning if $E$ and $\bar{E}$ are the energies of two different particles at time $t$, then $\omega = (E - \bar{E})/\hbar$ and at time $\bar{t}$, the difference in energy is given by the frequency $\bar{\omega}$.
The quantum noise correlation for the NQN junction for terminal $i = j = 1$ at vanishing frequency is given by~\cite{PhysRevB.53.16390, noise}:

\begin{widetext}
\begin{eqnarray}
Q_{11}^{NQN}(\omega = 0) &=& \frac{2e^2}{h} \int \sum_{p,q \in \{1,2\}} A_{p;l}(1,E) A_{q;k}(1,E) \left[f_p(E)(1 - f_q(E)) + f_q(E)(1 - f_p(E))\right] dE  \nonumber \\
&=& \frac{4e^2}{h} \biggl[ \int_{-\infty}^{\infty} F_{11\, \mathrm{eq}}^{NQN} \sum_{i \in \{ 1,2\}}\left(f_{ie}(1 - f_{ie})\right) dE + \int_{-\infty}^{\infty} F_{11\, \mathrm{sh}}^{NQN} (f_{1e} - f_{2e})^2 dE \biggr] = Q_{11\, \mathrm{eq}}^{NQN} + Q_{11\, \mathrm{sh}}^{NQN}.
\label{eqn:S_NIN}
\end{eqnarray}
\end{widetext}

Here, $A_{p; q}(1, E) = \delta_{1p} \delta_{1q} - s_{1p}^{\dagger} s_{1q}$ and $F_{11\, \mathrm{eq}}^{NQN} = \sum_{n} \mathcal{T}_{Q}(E)$ and $F_{11\, \mathrm{sh}}^{NQN} = \sum_{n}\mathcal{T}_{Q}(E)(1 - \mathcal{T}_{Q}(E))$ are derived in Appendix~\ref{App_ch_Qn} for multichannel case. The terms $Q_{11\, \mathrm{eq}}^{NQN}$ and $Q_{11\, \mathrm{sh}}^{NQN}$ correspond to {equilibrium} and shot noise contributions, respectively.

\subsubsection{NQS}

In this section, we discuss quantum noise in NQS junction and also decompose it into two physical quantities called quantum {equilibrium noise} and quantum shot noise.

Similarly for NQS junction, the current noise correlations between different quasiparticles (electrons and holes) has to be considered~\cite{PhysRevB.53.16390, martin2005course}. The noise correlation function is defined as
$Q_{ij}^{xy}(t - t') \equiv \langle \Delta I_i^x(t) \Delta I_j^y(t') + \Delta I_j^y(t') \Delta I_i^x(t) \rangle$,
where $\Delta I_i^x(t) = I_i^x(t) - \langle I_i^x(t) \rangle$ and $x, y \in \{e, h\}$. Its frequency-domain representation is
$\delta(\omega + \bar{\omega}) Q_{ij}^{xy}(\omega) \equiv \frac{1}{2\pi} \langle \Delta I_i^x(\omega) \Delta I_j^y(\bar{\omega}) + \Delta I_j^y(\bar{\omega}) \Delta I_i^x(\omega) \rangle$.
The zero-frequency noise in an NQS junction is given by:

\begin{widetext}
\begin{eqnarray}
Q_{11}^{NQS}(\omega = 0) &=&  \frac{2e^2}{h} \int \sum_{ \substack{p,q \in \{1, 2\} ,\\
\mu,\nu,\sigma,\rho \in \{e,h\}} } \mathrm{sgn}(\mu) sgn(\nu)  
A_{p,\sigma;q,\rho}(1 \mu) 
A_{q,\rho;p,\sigma}(1 \nu) 
({f}_{p \sigma} [1-{f}_{q \rho}]+(1-{f}_{p \sigma}) {f}_{q \rho}) \, dE, \nonumber \\
&=& Q_{11\, \mathrm{eq}}^{NQS} + Q_{11\, \mathrm{sh}}^{NQS},
\label{eqn:S_NIS}
\end{eqnarray}
\end{widetext}

where $\mathrm{sgn}(\mu) = +1$ for $x = e$ and $-1$ for $x = h$. Here, $A_{p,\sigma;q,\rho}(1 \mu) 
= \delta_{1 p} \delta_{1 q} \delta_{\mu \sigma} \delta_{\mu \rho} 
- s^{\mu \sigma *}_{1 p} s^{\mu \rho}_{1 q}$ and $V$ is the voltage bias applied. The {equilibrium} and shot noise contributions, $Q_{11\, \mathrm{eq}}^{NQS}$ and $Q_{11\, \mathrm{sh}}^{NQS}$, are derived in Appendix~\ref{App_ch_Qn} and for multichannel case ($n>0$), they are given as,

\begin{widetext}
\begin{equation}
    \begin{split}
        Q_{11eq}^{NQS} &= \frac{4e^2}{h} \int_{-\infty}^{\infty} dE \sum_{n}\bigg[(1 - \mathcal{R}^B(E) + \mathcal{R}^A(E))^2 + \mathcal{R}^A(-E) \mathcal{R}^B(E) + \mathcal{R}^A(E) \mathcal{R}^B(-E) + \mathcal{R}^B(E)(\mathcal{T}^C(E) + \mathcal{T}^D(-E)) \\& + \mathcal{R}^A(E)(\mathcal{T}^D(E) + \mathcal{T}^C(-E) + 2 \mathcal{R}^A(E) \mathcal{R}^B(E))\bigg]   f_{1e}(1-f_{1e}) + \frac{4e^2}{h} \int_{-\infty}^{\infty} dE \sum_{n}\bigg[\mathcal{T}^C(E)^2 + \mathcal{T}^D(E)^2 \\& + 2 \mathcal{T}^C(E) \mathcal{T}^D(-E) + \mathcal{R}^B(E) (\mathcal{T}^C(E) + \mathcal{T}^D(-E)) + \mathcal{R}^A(E)(\mathcal{T}^C(-E) + \mathcal{T}^D(E))\bigg]f_{2e}(1-f_{2e}),\\
        Q_{11sh}^{NQS} &= \frac{4e^2}{h} \int_{-\infty}^{\infty}dE \sum_{n}\bigg[\mathcal{R}^B(E) (\mathcal{T}^C(E) + \mathcal{T}^D(-E)) + \mathcal{R}^A(E) (\mathcal{T}^C(-E) + \mathcal{T}^D(E)) + 2 \mathcal{R}^A(E) \mathcal{R}^B(E) \\&+ 2 Re(s_{11}^{ee}(E) s_{11}^{ee * }(-E) s_{11}^{he * }(E) s_{11}^{he  }(-E)) \bigg]  (f_{1e} - f_{2e})^2 + \frac{4e^2}{h} \int_{-\infty}^{\infty}dE \sum_{n}\bigg[\mathcal{R}^A(E) \mathcal{R}^B(E)\\& -  Re(s_{11}^{ee}(E) s_{11}^{ee * }(-E) s_{11}^{he * }(E) s_{11}^{he  }(-E)) \bigg](f_{1e} - f_{1h})^2.
    \end{split}
     \label{chqnoiseNQS}
\end{equation}
   
\end{widetext}

\subsection{$\Delta_T$ noise}

In this section, we provide a detailed explanation of $\Delta_T$ noise in both NQN and NQS junctions, along with the methodology used for its computation.

This subsection discusses the $\Delta_T$ noise in an NQN junction.

\subsubsection{NQN}

In an NQN junction, the $\Delta_T$
noise corresponds to a shot noise–type contribution that remains finite when the charge current vanishes in the presence of a nonzero {thermovoltage}. We denote this quantity as $\Delta_T^{NQN}$, and it is calculated at the {thermovoltage} $V_{th}^{NQN}$, as defined in Sec.~\ref{theory} A 1. Hence, $\Delta_T^{NQN}$ corresponds to $Q_{11sh}^{NQN}$ in Eq.~(\ref{eqn:S_NIN}), evaluated at $V_{th}^{NQN}$. The explicit expression for $\Delta_T^{NQN}$ is:

\begin{equation}
    \Delta_T^{NQN} = \frac{4e^2}{h} \int_{-\infty}^{\infty} dE\, F_{11sh}^{NQN} (f_{1e} - f_{2e})^2,
\end{equation}

where $f_{1e}$ and $f_{2e}$ are evaluated at the finite {thermovoltage} $V_{th}^{NQN}$.

We also look at normalized $\Delta_T$ noise, which we define as the ratio of the $\Delta_T$ noise and conductance. For the NQN junction, normalized $\Delta_T$ noise is denoted as $\bar{\Delta}_T^{NQN}$, which is equal to $\frac{\Delta_T^{NQN}}{G_{NQN}}$, where the expression for $G_{NQN}$ is given in Eq. (\ref{eqn10}). The normalized $\Delta_T$ noise plays a central role, as it is similar to Fano factor, which is basically the effective charge transported
through the junction. Therefore, $\bar{\Delta}_T^{NQN}$ is basically a measure of the effective charge transported through the NQN junction.

\subsubsection{NQS}

In this subsection, we examine the behavior of $\Delta_T$ noise associated with the NQS junction. $\Delta_T$ noise in an NQS junction refers to the quantum shot noise-type contribution, evaluated at zero current and under a finite {thermovoltage}. It is denoted as $\Delta_T^{NQS}$ and is evaluated at the {thermovoltage} $V_{th}^{NQS}$, as defined in Sec.~\ref{theory} A 2. Therefore, $\Delta_T^{NQS}$ corresponds to $Q_{11sh}^{NQS}$, given in Eq.~(\ref{chqnoiseNQS}), evaluated at $V_{th}^{NQS}$. The expression is:

\begin{equation}
\begin{split}
   & \Delta_T^{NQS} = \frac{4e^2}{h} \int_{-\infty}^{\infty} dE\, \sum_{n}\bigg[\mathcal{R}^B(E)(\mathcal{T}^C(E) + \mathcal{T}^D(-E)) + \mathcal{R}^A(E)\\&(\mathcal{T}^C(-E)  + \mathcal{T}^D(E)) + 2\mathcal{R}^A(E)\mathcal{R}^B(E) + 2\,\text{Re}(s_{11}^{ee}(E) s_{11}^{ee * }(-E)\\& s_{11}^{he * }(E) s_{11}^{he  }(-E)) \bigg]  (f_{1e} - f_{2e})^2 + \frac{4e^2}{h} \int_{-\infty}^{\infty} dE\, \sum_{n}\bigg[\mathcal{R}^A(E)\mathcal{R}^B(E) \\& - \text{Re}(s_{11}^{ee}(E) s_{11}^{ee * }(-E) s_{11}^{he * }(E) s_{11}^{he  }(-E)) \bigg](f_{1e} - f_{1h})^2.
\end{split}
\end{equation}

Here, $f_{1e}$, $f_{1h}$, and $f_{2e}$ denote the Fermi-Dirac distributions for electrons and holes in terminal 1 and for electrons in terminal 2, respectively, all evaluated at the finite {thermovoltage} $V_{th}^{NQS}$.

Similar to NQN junction, we also look at normalized $\Delta_T$ noise in NQS junction, which is denoted as $\bar{\Delta}_T^{NQS}$, which is equal to $\frac{\Delta_T^{NQS}}{G_{NQS}}$, where the expression for $G_{NQS}$ is given in Eq. (\ref{eqn11}). $\bar{\Delta}_T^{NQS}$ is a measure of the effective charge transported through the NQS junction.

\section{Results and Discussion}
\label{results}

{The key results of this work are reported in this section, which is divided into two parts.  First, we focus on the characteristics of ($V_{th}^{NQN}$) {thermovoltage} in NQN as well as ($V_{th}^{NQS}$) {thermovoltage} in NQS setup and further we calculate the $\Delta_T$ noise as well as quantum noise in NQN and NQS junctions.}

\subsection{{Thermovoltage}}

\subsubsection{NQN}

In Fig.~\ref{fig5}(a), we illustrate the behavior of the {thermovoltage}, denoted as \( V_{th}^{NQN} \), as a function of the dimensionless parameter, i.e. normalized Fermi energy \( \frac{E_F - V_0}{\hbar \omega_x} \), under the condition of zero net charge current. The analysis is carried out at various average temperatures: \( T =0.01K, 0.1K,  1.0K, 3.0K\), and \( 5.0\,K \), and for four distinct transport channels. For the NQN junction, \( V_{th}^{NQN} \) oscillates with increasing \( \frac{E_F - V_0}{\hbar \omega_x} \). This behavior stems from the oscillating behavior of the Seebeck coefficient $S_{NQN}$ in presence of QPC, see Fig. \ref{fig5}(c). Here, we ignore the range of $\frac{E_F - V_0}{\hbar \omega_x}$ between 0 to 1 as there are no QPC energy levels in that range.

\subsubsection{NQS}

\begin{widetext}

\begin{figure}[H]
\centering
\includegraphics[scale=0.30]{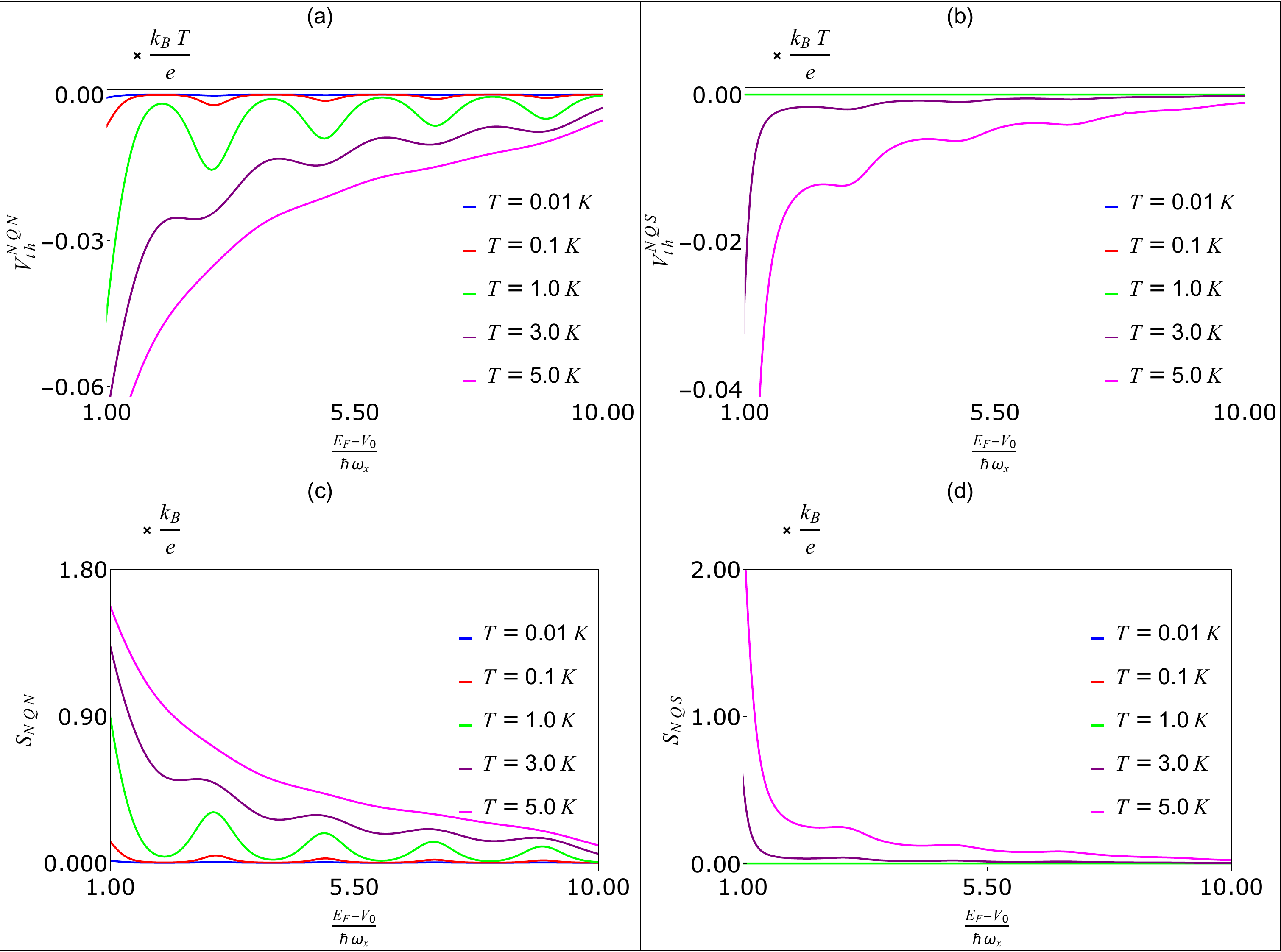}
\caption{{thermovoltage} in (a) NQN ($V^{NQN}_{th}$) and (b) NQS ($V_{th}^{NQS}$) (in units of $\frac{k_B T}{e}$), Seebeck coefficient in (c) NQN ($S_{NQN}$) and (d) NQS ($S_{NQS}$) (in units of $\frac{k_B}{e}$)  vs. normalized Fermi energy $\frac{E_F - V_0}{\hbar \omega_x}$. The parameters taken are $n = 4$, $ \hbar \omega_y = 1.365 meV$, $\hbar \omega_x =0.5 \hbar \omega_y$ and $\Delta T = 0.05 T$, where $T_C = 9K$. {The curves corresponding to the temperatures $T = 0.01K,0.1K$ in (b) and (d) nearly overlap and are therefore difficult to distinguish on the scale of the figure}.}
\label{fig5}
\end{figure}
\end{widetext}

In Fig.~\ref{fig5}(b), we show the behavior of the {thermovoltage} \( V_{th}^{NQS} \), evaluated under the same condition of zero net charge current. Below $T = 3.0K$, i.e, at $T = 0.01K, 0.1K$ and $1.0K$, thermal excitations are weak, and the corresponding excitation energies remain below the superconducting gap. In this low-energy regime, electron–hole (e–h) symmetry is effectively preserved because the transmission probabilities \( \mathcal{R}^A(E) \) and \( \mathcal{R}^B(E) \) are symmetric with respect to energy \( E \), resulting in a vanishing {thermovoltage}, which is what is seen in Fig. \ref{fig5}(b). As the temperature increases, thermal excitations become stronger, enabling excitation energies to exceed the energy scale of the barrier, thereby breaking e–h symmetry and inducing a finite {thermovoltage}. However, as the temperature increases to \( T = 3.0\,K,\ 5.0\,K\), thermal excitations become sufficiently strong to populate quasiparticle states above the superconducting gap, thereby breaking e–h symmetry and giving rise to a finite {thermovoltage}. In contrast to the oscillating trend observed in the NQN case, \( V_{th}^{NQS} \) also exhibits oscillating dependence on \( \frac{E_F - V_0}{\hbar \omega_x} \). However, the magnitude of $V_{th}^{NQS}$ at $T = 3.00K$ is small and not visible to the naked eye. However, as we increase the temperature upto $T = 5.00K$, $V_{th}^{NQS}$ shows prominent oscillations. This comes as a consequence of the oscillation of the Seebeck coefficient $S_{NQS}$ as a function of $\frac{E_F - V_0}{\hbar \omega_x}$, see Fig. \ref{fig5}(d).

\begingroup

The key point is that the thermoelectric response is governed by the symmetry of the net transmission probability of the NQS junction, $
\mathcal{T}_{NQS}(E)=1+\mathcal{R}^{A}(E)-\mathcal{R}^{B}(E).$ This directly determines the behavior of the thermoelectric coefficient, $
\Pi_{NQS}=\frac{2e}{hT}\int_{-\infty}^{\infty} dE \sum_n \mathcal{T}_{NQS}(E)E\Big(-\frac{\partial f}{\partial E}\Big),$
which is responsible for generating the thermovoltage through the zero current condition, $
V_{\mathrm{th}}^{NQS}=-\frac{\Pi_{NQS}}{G_{NQS}}\Delta T.$

At low temperatures, $k_BT\ll\Delta_0$, the thermal broadening of the Fermi distribution originating from $\Big(-\frac{\partial f}{\partial E}\Big)$ is very narrow. The function $\Big(-\frac{\partial f}{\partial E}\Big)$ is always centered at $E=0$, while its width is determined by the thermal energy $k_BT$. Since the center of this thermal window lies inside the superconducting gap and its width is much smaller than $\Delta_0$, the integrand receives appreciable contributions almost entirely from subgap energies ($|E|<\Delta_0$), whereas contributions from above-gap quasiparticle states are suppressed. Consequently, transport is dominated by the subgap scattering processes. In this energy regime, $\mathcal{T}_{NQS}(E)$ is symmetric with $E$, and therefore the thermoelectric coefficient $\Pi_{NQS}$ is zero, leading to a vanishing thermovoltage.

As the temperature increases, the thermal broadening increases proportionally to $k_BT$, thereby enlarging the energy window over which $\Big(-\frac{\partial f}{\partial E}\Big)$ has significant weight. Although the center of the distribution remains fixed at $E = 0$, its broadening extend beyond the superconducting gap, allowing above-gap quasiparticle states ($|E|>\Delta_0$) to participate in transport. In this energy regime, $\mathcal{T}_{NQS}(E)$ is asymmetric with respect to energy. This finite electron-hole asymmetry gives rise to a nonzero thermoelectric coefficient $\Pi_{NQS}$, which in turn generates a finite thermovoltage through the zero-current condition.

Therefore, at low temperatures transport is dominated by the subgap regime, where the effective transmission of the NQS junction remains nearly electron-hole symmetric. However, when thermally excited quasiparticles above the superconducting gap begin to participate in transport, the electron-hole asymmetry leads to finite thermovoltage.

\endgroup

%\subsection{$\Delta_T$ charge noise in NQN and NQS junction ($V_1 - V_2 = V_{th}, T_1 - T_2 = -\Delta T)$)}
%\label{results_DTnoise}

\subsection{$\Delta_T$ noise}

\subsubsection{NQN}

\begin{widetext}

\begin{figure}[H]
\centering
\includegraphics[scale=0.30]{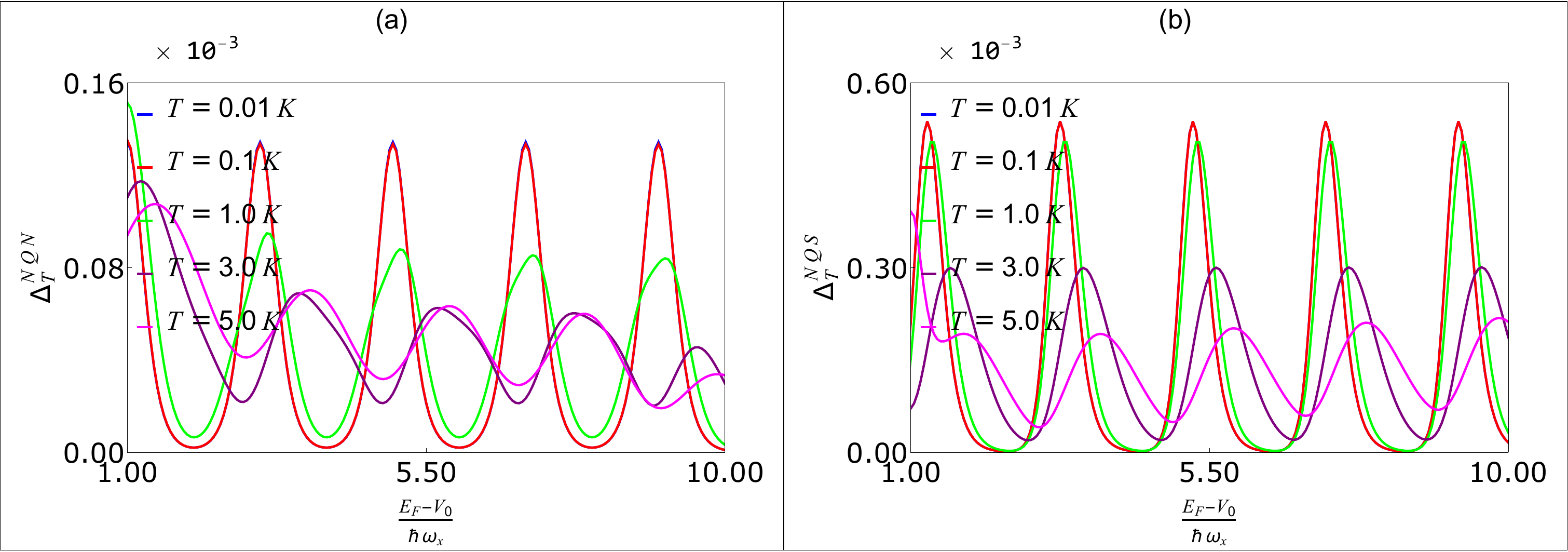}
\caption{Charge $\Delta_T$ noise in (a) NQN ($\Delta^{NQN}_{T}$) and (b) NQS ($\Delta_{T}^{NQS}$) (in units of $\frac{4e^2}{h} k_B T$) vs. normalized Fermi energy $\frac{E_F - V_0}{\hbar \omega_x}$. The parameters taken are $n = 4$, $ \hbar \omega_y = 1.365 meV$, $\hbar \omega_x =0.5 \hbar \omega_y$ and $\Delta T = 0.05 T$, where $T_C = 9K$. {The curves corresponding to the temperatures $T = 0.01K,0.1K$ nearly overlap and are therefore difficult to distinguish on the scale of the figure}.}
\label{fig6}
\end{figure}
\end{widetext}

In Fig.~\ref{fig6}(a), we present the behavior of the charge \( \Delta_T \) noise for the NQN junction, denoted by \( \Delta_T^{NQN} \), as a function of the dimensionless parameter \( \frac{E_F - V_0}{\hbar \omega_x} \). The data exhibits clear oscillations in \( \Delta_T^{NQN} \) as \( \frac{E_F - V_0}{\hbar \omega_x} \) varies. Notably, \( \Delta_T^{NQN} \) tends to approach zero in regions where the charge conductance \( G_{NQN} \) displays plateau behavior, as shown in Fig.~\ref{fig3}(a). In contrast, peaks in \( \Delta_T^{NQN} \) are observed near the transitions between conductance plateaus. This behavior is physically intuitive: in regions of perfect transmission (transparency), shot noise is suppressed, whereas across transitions, where the transmission probability deviates from unity, the shot noise and, hence, \( \Delta_T \) noise are enhanced. As the temperature increases, the amplitude of oscillations in \( \Delta_T^{NQN} \) starts to reduce due to thermal broadening and the influence of finite {thermovoltage}.

\subsubsection{NQS}

In Fig.~\ref{fig6}(b), we plot the corresponding \( \Delta_T \) noise for the NQS junction, denoted by \( \Delta_T^{NQS} \), evaluated under the same conditions. While both NQN and NQS junctions exhibit qualitatively similar oscillatory behavior, reflecting the thermal origin of the noise. As temperature increases, the contribution from the finite {thermovoltage} becomes more prominent, leading to a gradual reduction in both \( \Delta_T^{NQN} \) and \( \Delta_T^{NQS} \) from their low-temperature values.

\subsection{Quantum noise}

\begingroup

\subsubsection{NQN}

We now examine the quantum noise measured at a finite voltage bias, where a nonzero charge current flows through the junction. In this regime, the nature of the noise differs significantly from the $\Delta_T$ noise discussed previously. The behavior of the quantum noise $Q_{11}^{\mathrm{NQN}}$ is shown in Fig.~\ref{fig7}(a) as a function of the dimensionless parameter $\frac{E_F - V_0}{\hbar \omega_x}$ for several temperatures.

As seen in the figure, $Q_{11}^{\mathrm{NQN}}$ exhibits pronounced step-like structures across the entire temperature range considered. These steps originate from the energy-dependent transmission of the quantum point contact (QPC). As the parameter $\frac{E_F - V_0}{\hbar \omega_x}$ is varied, successive transverse modes of the QPC become available for transport. Each newly opened channel contributes an additional unit of transmission, producing discrete increases in the transport properties of the system. Consequently, the quantum noise inherits a step-like dependence that closely resembles the well-known quantized conductance behavior of a QPC.

It is important to emphasize that, in contrast to the $\Delta_T$ noise discussed earlier, the quantum noise does not display oscillatory features. This difference arises from the distinct physical origins of the two quantities. The quantum noise measured at finite voltage bias contains contributions from both {equilibrium noise} and shot noise–like fluctuations. In the present parameter regime, the {equilibrium noise} component dominates the shot noise contribution. Since {equilibrium noise} is directly related to the thermal conductance of the junction, the resulting quantum noise reflects the same step-like structure that characterizes the conductance of a QPC. As a result, the overall behavior of $Q_{11}^{\mathrm{NQN}}$ closely follows the quantized transmission profile of the contact rather than exhibiting oscillatory variations.

\subsubsection{NQS}

We now turn to the quantum noise in the NQS junction under finite voltage bias. The corresponding results for $Q_{11}^{\mathrm{NQS}}$ are presented in Fig.~\ref{fig7}(b) as a function of $\frac{E_F - V_0}{\hbar \omega_x}$ for different temperatures. The overall behavior remains qualitatively similar to that observed in the NQN configuration.

In particular, the quantum noise again displays clear step-like features as the QPC parameter is varied. These steps arise from the same physical mechanism described in the NQN case, namely the sequential opening of transverse transport channels in the quantum point contact. As additional channels become available, the transmission probability increases discretely, and the quantum noise correspondingly increases in a stepwise manner. The persistence of these steps across a wide temperature range indicates that the dominant contribution to the noise is governed by the underlying channel structure of the QPC.

\begin{widetext}

\begin{figure}[H]
\centering
\includegraphics[scale=0.30]{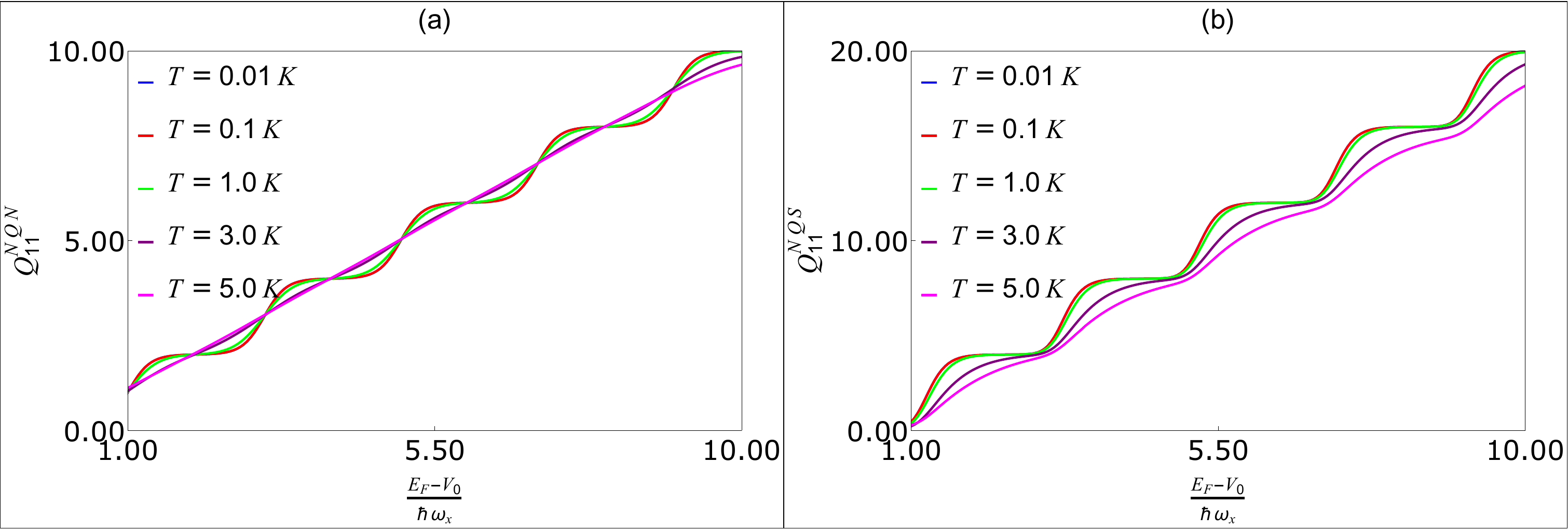}
\caption{Quantum noise in (a) NQN ($Q^{NQN}_{11}$) and (b) NQS ($Q_{11}^{NQS}$) (in units of $\frac{4e^2}{h} k_B T$) vs. normalized Fermi energy $\frac{E_F - V_0}{\hbar \omega_x}$. The parameters taken are $eV = 0.1k_BT$, $n = 4$, $ \hbar \omega_y = 1.365 meV$, $\hbar \omega_x =0.5 \hbar \omega_y$ and $eV=0.1k_B T$, where $T_C = 9K$. {The curves corresponding to the temperatures $T = 0.01K,0.1K$ nearly overlap and are therefore difficult to distinguish on the scale of the figure}.}
\label{fig7}
\end{figure}
\end{widetext}

A notable difference between the two junctions is the magnitude of the quantum noise. We find that the quantum noise in the NQS configuration is approximately twice as large as that in the NQN junction. This enhancement originates from the presence of superconductivity and the associated Andreev reflection processes at the interface. Through Andreev reflection, an incident electron from the normal region is retroreflected as a hole while a Cooper pair is transferred into the superconductor. This process effectively doubles the charge transfer involved in the transport process, which leads to an enhanced noise magnitude in comparison with the purely normal junction.

Similar to the NQN case, the total quantum noise in the NQS junction consists of both {equilibrium noise} and shot noise–like contributions. In the parameter regime considered here, the {equilibrium noise} component remains dominant. Consequently, the overall noise profile primarily reflects the thermal conductance of the junction and therefore retains the characteristic step-like features associated with quantized transport through the quantum point contact.

\endgroup

\section{Analysis}
\label{analysis}

{This section analyzes our results of conductance, Seebeck coefficient, quantum noise and $\Delta_T$ noise.}

\begingroup

\subsection{Conductance}

Figure~\ref{fig10}(a) presents a comparison of the conductance in the NQS and NQN junctions through the ratio 
$\frac{G_{NQS}}{G_{NQN}}$ as a function of the dimensionless parameter 
$\frac{E_F - V_0}{\hbar \omega_x}$ in the regime 
$1 \leq \frac{E_F - V_0}{\hbar \omega_x} \leq 10$, where electron--hole (e--h) symmetry is strongly broken by the presence of the quantum point contact (QPC). In this region, the parameter range crosses the QPC energy levels 
$E_n = \left(n+\frac{1}{2}\right)\hbar\omega_y$ for $n=0\text{--}4$, wherein $\omega_y = 2\omega_x$, $E_F$ denotes the Fermi energy, $V_0$ the QPC confinement potential, and $\omega_x$ the characteristic frequency of the QPC. 
The numerical values corresponding to this behavior are summarized in Table~\ref{Table3}. 
We observe that the upper bound of this ratio approaches the value 2, which is expected due to perfect Andreev reflection occurring at the interface in the NQS junction. 
At ultra–low temperatures ($T = 0.01\,K$), the conductance ratio remains very close to this upper bound. 
However, as the temperature increases, the ratio slowly decreases from this upper bound. 
This reduction originates from e-h symmetry breaking in these regimes at finite temperatures, which gives rise to {thermovoltage}.

\begin{widetext}

\begin{figure}[H]
\centering
\includegraphics[scale=0.30]{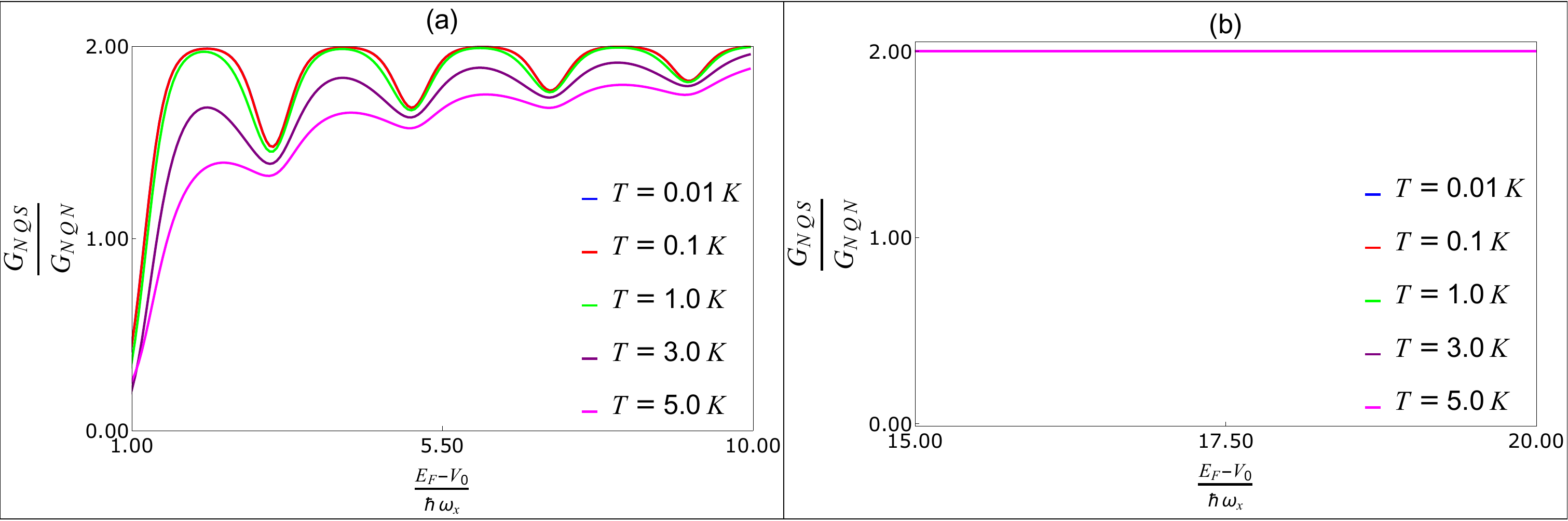}
\caption{Ratio of conductance $\frac{G_{NQS}}{G_{NQN}}$ for (a) e-h asymmetric, (b) e-h symmetric case vs. $\frac{E_F - V_0}{\hbar \omega_x}$. Parameters: $n = 4$, $ \hbar \omega_y = 1.365 meV$, $\hbar \omega_x =0.5 \hbar \omega_y$ and $\Delta T = 0.05 T$, where $T_C = 9K$.  {The curves corresponding to the temperatures $T = 0.01K,0.1K$ nearly overlap in (a) and are therefore difficult to distinguish on the scale of the figure. All curves in (b) nearly overlap and are difficult to distinguish}.}
\label{fig10}
\end{figure}
\end{widetext}

\begin{figure}[H]
\centering
\includegraphics[scale=0.25]{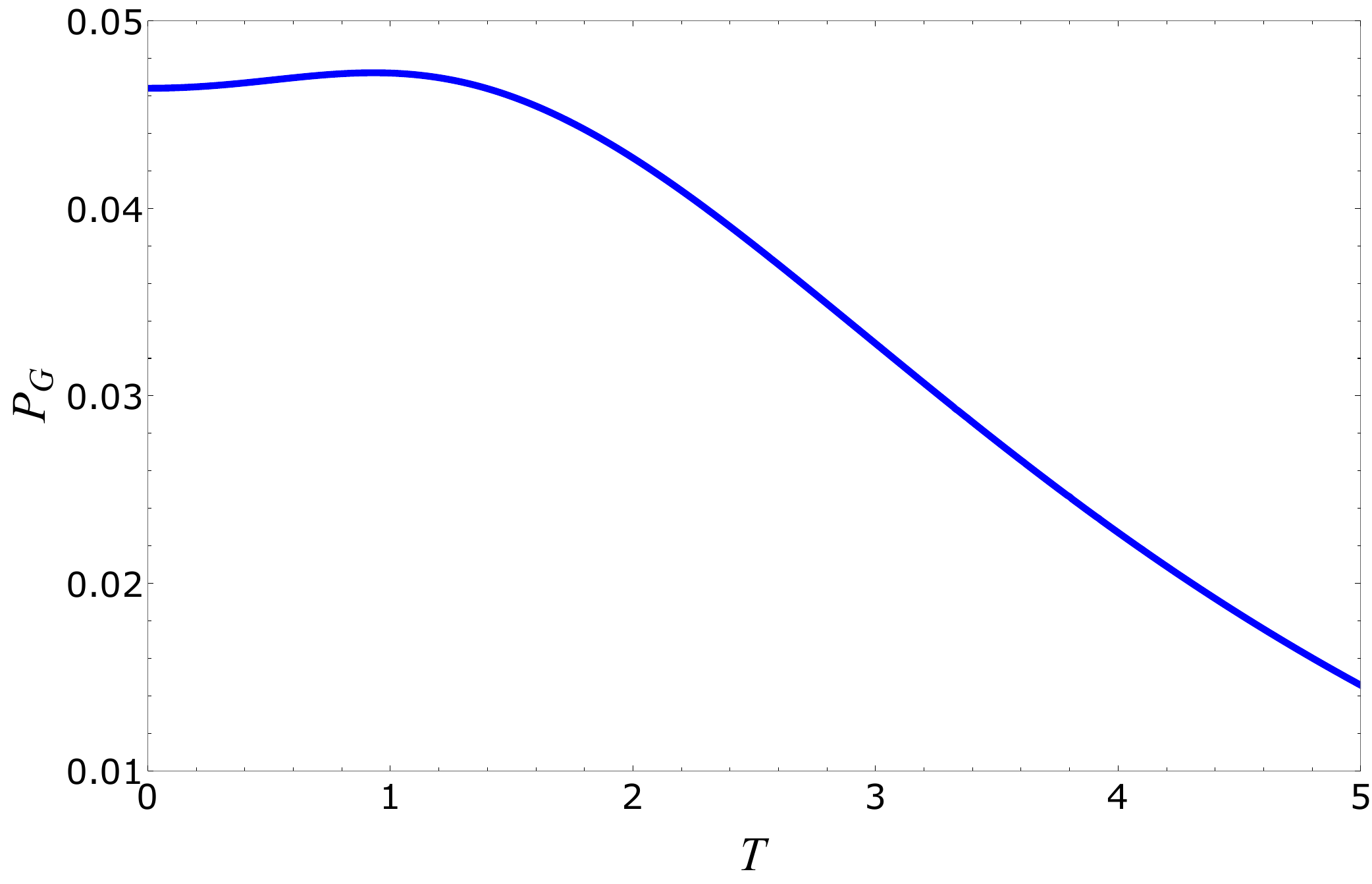}
\caption{Visibility for conductance ($P_G$) vs. $T$. Parameters: $n = 4$, $ \hbar \omega_y = 1.365 meV$, $\hbar \omega_x =0.5 \hbar \omega_y$ and $\Delta T = 0.05 T$, where $T_C = 9K$.}
\label{figPG}
\end{figure}

To provide a proper reference point for this behavior, we also examine the regime 
$\frac{E_F - V_0}{\hbar \omega_x} \gg 10$, where the QPC becomes nearly transparent and the transport effectively restores electron--hole symmetry. 
In this symmetric regime the ratio $\frac{G_{NQS}}{G_{NQN}}$ remains exactly equal to 2 for all temperatures considered, as listed in Table~\ref{Table3}. 
This visibility of the ratio: $P_G = \frac{max\left(\frac{G_{NQS}}{G_{NQN}}\right)-min\left(\frac{G_{NQS}}{G_{NQN}}\right)}{max\left(\frac{G_{NQS}}{G_{NQN}}\right)+min\left(\frac{G_{NQS}}{G_{NQN}}\right)}$ is not striking enough to be really useful for detecting nature of quantum transport or e-h symmetry breaking as compared to $\Delta_T$ noise as we show later (see, Fig. \ref{figPG}), visibility also reduces with increasing energies.

Conductance primarily probes the average charge transport across the junction. In superconducting hybrid structures, transport at low energies is dominated by Andreev reflection, where an incident electron from the normal region is reflected as a hole while a Cooper pair is transferred into the superconductor. This process effectively doubles the transferred charge and leads to the characteristic enhancement of conductance in superconducting junctions. Consequently, the ratio $\frac{G_{NQS}}{G_{NQN}}$ approaches the universal value of $2$ in the transparent limit. When the quantum point contact breaks electron--hole symmetry, a finite thermoelectric response emerges in the system. However, since conductance reflects only the average transport properties, it remains only weakly sensitive to the underlying electron--hole asymmetry introduced by the QPC.

\endgroup

\begingroup

\subsection{Seebeck Coefficient}
Fig.~\ref{figS} shows the ratio of the Seebeck coefficients of the NQS and NQN junctions, 
$\frac{S_{NQS}}{S_{NQN}}$, plotted as a function of the dimensionless parameter 
$\frac{E_F - V_0}{\hbar \omega_x}$ for several temperatures. 
The Seebeck coefficient characterizes the thermoelectric response of the system, namely the voltage generated in response to a temperature gradient across the junction, and therefore provides information about electron–hole (e–h) asymmetry in transport.

\begin{figure}[H]
\centering
\includegraphics[scale=0.25]{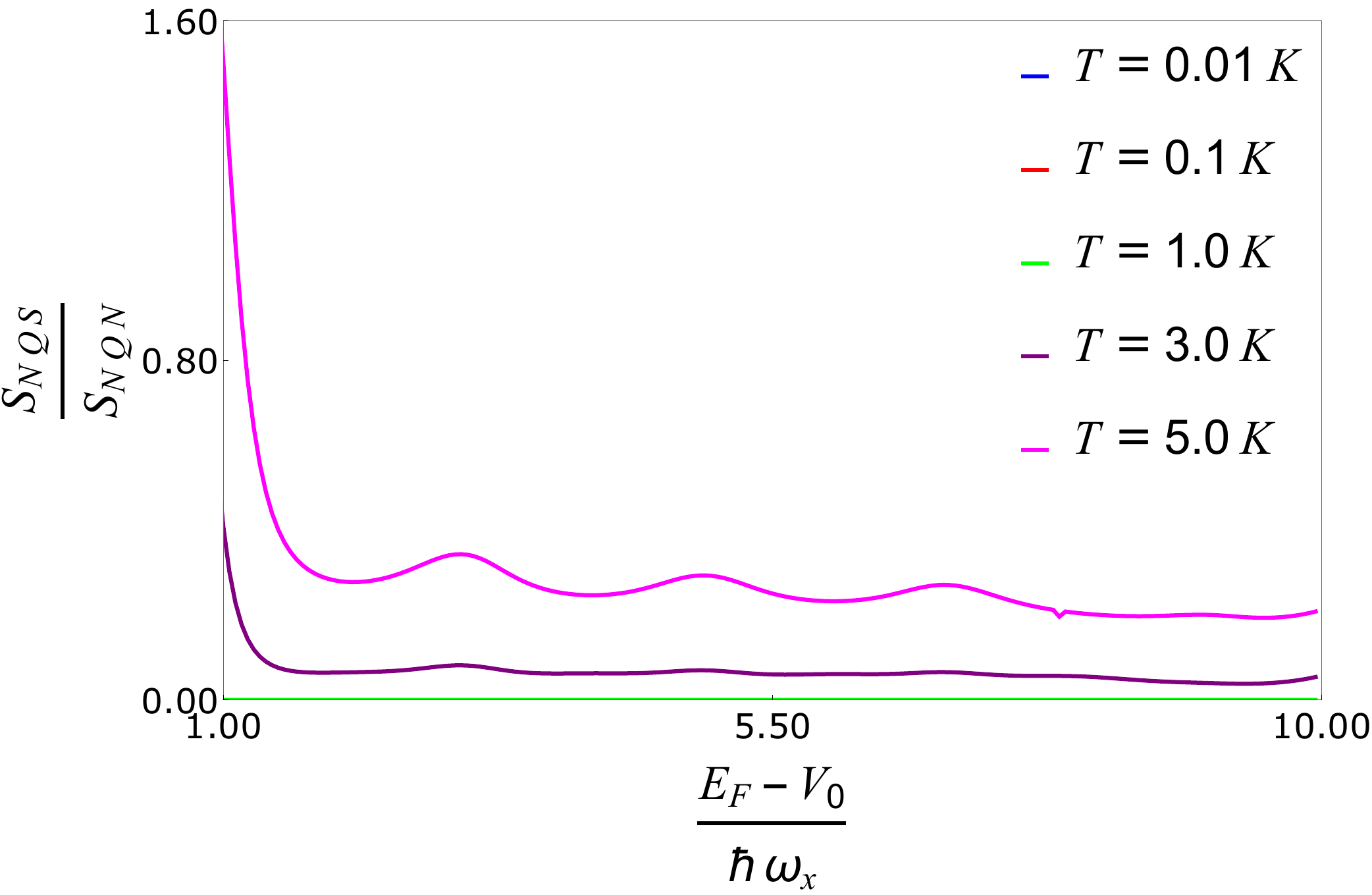}
\caption{Ratio of Seebeck coefficient $\frac{S_{NQS}}{S_{NQN}}$ vs. $\frac{E_F - V_0}{\hbar \omega_x}$. Parameters: $n = 4$, $ \hbar \omega_y = 1.365 meV$, $\hbar \omega_x =0.5 \hbar \omega_y$ and $\Delta T = 0.05 T$, where $T_C = 9K$. {The curves corresponding to the temperatures $T = 0.01K,0.1K$ nearly overlap and are therefore difficult to distinguish on the scale of the figure}.}
\label{figS}
\end{figure}

\begin{figure}[H]
\centering
\includegraphics[scale=0.25]{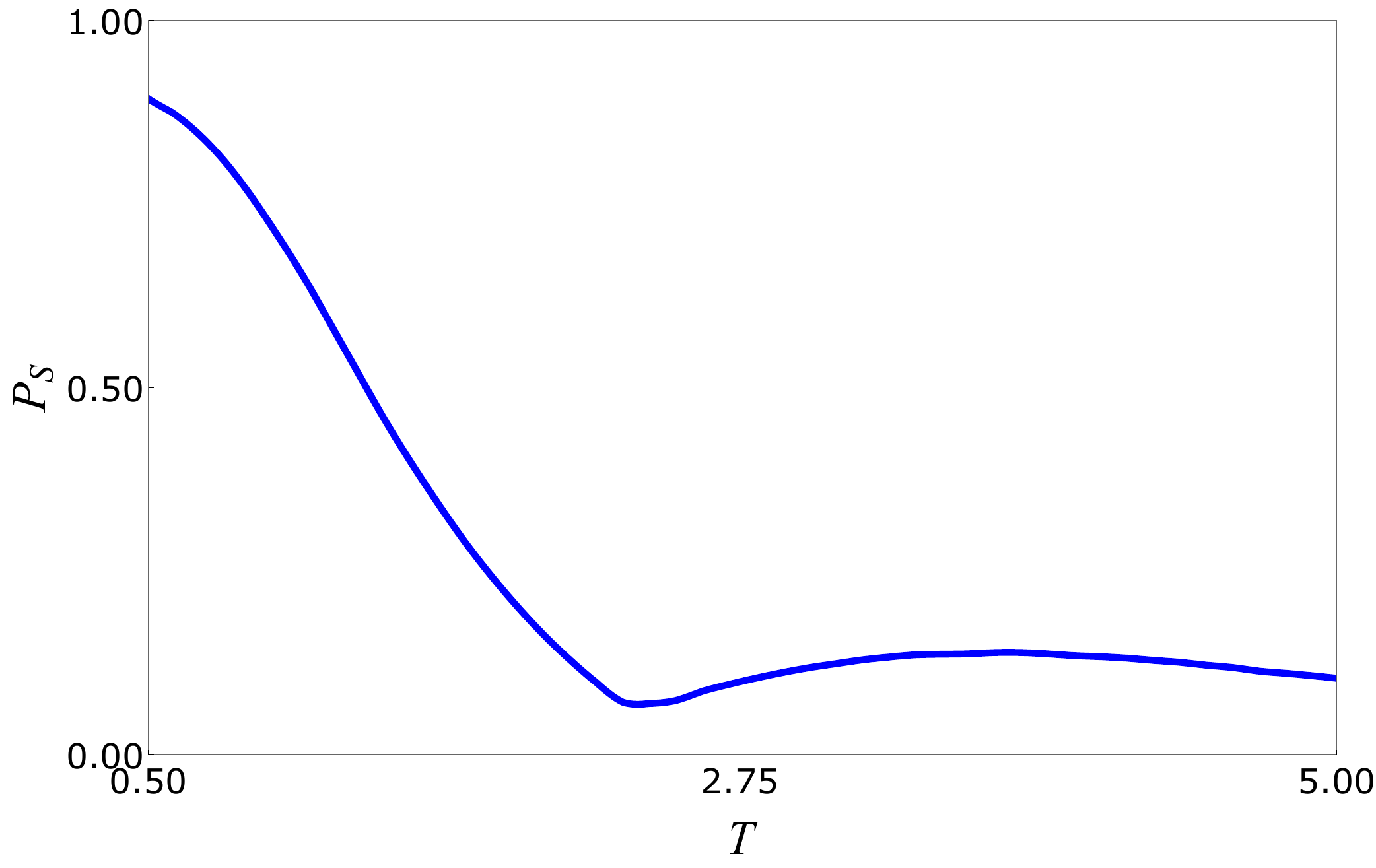}
\caption{Visibility for Seebeck coefficient ($P_S$) vs. $T$. Parameters: $n = 4$, $ \hbar \omega_y = 1.365 meV$, $\hbar \omega_x =0.5 \hbar \omega_y$ and $\Delta T = 0.05 T$, where $T_C = 9K$.}
\label{figPS}
\end{figure}

We observe that the ratio of the Seebeck coefficients remains extremely small at lower temperatures ($T = 0.01\,K$, $0.1\,K$, and $1\,K$), indicating that the thermoelectric response is strongly suppressed in this regime. As the temperature increases, however, the ratio begins to exhibit oscillatory behavior around a small but finite value. For instance, at $T = 3\,K$ the ratio oscillates around approximately $0.005$, while at $T = 5\,K$ the oscillations occur around approximately $0.03$, as summarized in Table~\ref{Table3}. These oscillations originate from the interplay between the energy-dependent transmission of the QPC and the electron–hole asymmetry introduced in the junction.

Although the Seebeck coefficient is directly related to electron–hole symmetry breaking and therefore provides information about thermoelectric effects induced by the QPC, its magnitude remains relatively small over the temperature range considered here. Consequently, while it reflects the presence of electron–hole asymmetry, the weak amplitude of the Seebeck response limits its ability to serve as a clear and robust probe of the underlying symmetry breaking in superconducting hybrid junctions. However, we see that the visibility: $P_S = \frac{max\left(\frac{S_{NQS}}{S_{NQN}}\right)-min\left(\frac{S_{NQS}}{S_{NQN}}\right)}{max\left(\frac{S_{NQS}}{S_{NQN}}\right)+min\left(\frac{S_{NQS}}{S_{NQN}}\right)}$ is quite significant, see Fig. \ref{figPS}. However, we restrict the plot of $P_S$ for $T$ from 0.50 to 1 because from $T = 0$ to 0.50, the Seebeck coefficient is always zero.

\endgroup

\begingroup

\subsection{Quantum noise}

Figure~\ref{fig11} shows the ratio of the quantum noise in the NQS and NQN junctions, 
$\frac{Q_{11}^{\mathrm{NQS}}}{Q_{11}^{\mathrm{NQN}}}$, plotted as a function of the dimensionless parameter 
$\frac{E_F - V_0}{\hbar \omega_x}$ for several temperatures. 
The corresponding numerical values are also summarized in Table~\ref{Table3}. 
At the lowest temperature considered ($T = 0.01\,K$), the ratio approaches an upper bound of 2. 
This behavior is consistent with the presence of superconductivity in the NQS junction, where Andreev reflection processes effectively enhance the noise relative to the purely normal NQN junction.

As the temperature increases, however, the ratio gradually decreases from this upper bound. 
This reduction arises from the combined effects of thermal excitations and electron--hole symmetry breaking induced by the QPC. 
At higher temperatures, thermally excited quasiparticles populate states above the superconducting gap, thereby weakening the relative contribution of Andreev reflection to transport. 
Consequently, the enhancement of quantum noise in the superconducting junction becomes less pronounced, leading to a decrease in the ratio 
$\frac{Q_{11}^{\mathrm{NQS}}}{Q_{11}^\mathrm{NQN}}$ with increasing temperature.

To place this behavior in the proper context, we also examine a regime where electron--hole symmetry is effectively preserved. 
Specifically, we consider the region 
$\frac{E_F - V_0}{\hbar \omega_x} \gg 10$, where the QPC becomes nearly transparent and the transport is electron--hole symmetric. 
The corresponding results are shown in Fig.~\ref{fig11}(b) and summarized in Table~\ref{Table3}. 
In this regime, the ratio $\frac{Q_{11}^{\mathrm{NQS}}}{Q_{11}^{\mathrm{NQN}}}$ remains exactly at 2 for all temperatures considered. 
This temperature–independent behavior reflects the robustness of the symmetric transport regime, where Andreev reflection processes enhance the noise by a factor of two compared to the normal junction.

\begin{widetext}

\begin{figure}[H]
\centering
\includegraphics[scale=0.30]{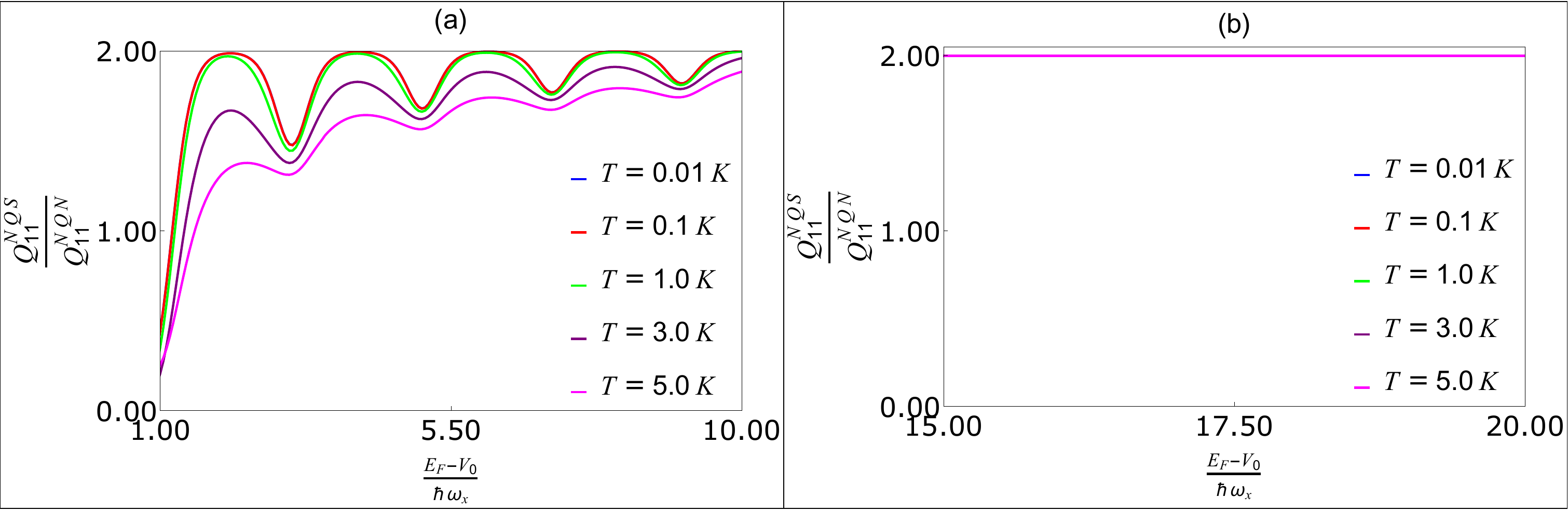}
\caption{Ratio of quantum noise $\frac{Q_{11}^{NQS}}{Q_{11}^{NQN}}$ for (a) e-h asymmetric, (b) e-h symmetric case vs. $\frac{E_F - V_0}{\hbar \omega_x}$. Parameters: $n = 4$, $ \hbar \omega_y = 1.365 meV$, $\hbar \omega_x =0.5 \hbar \omega_y$ and $eV=0.1k_B T$, where $T_C = 9K$. {The curves corresponding to the temperatures $T = 0.01K,0.1K$ nearly overlap in (a) and are therefore difficult to distinguish on the scale of the figure. All curves in (b) nearly overlap and are difficult to distinguish.}}
\label{fig11}
\end{figure}

\end{widetext}

\begin{figure}[H]
\centering
\includegraphics[scale=0.25]{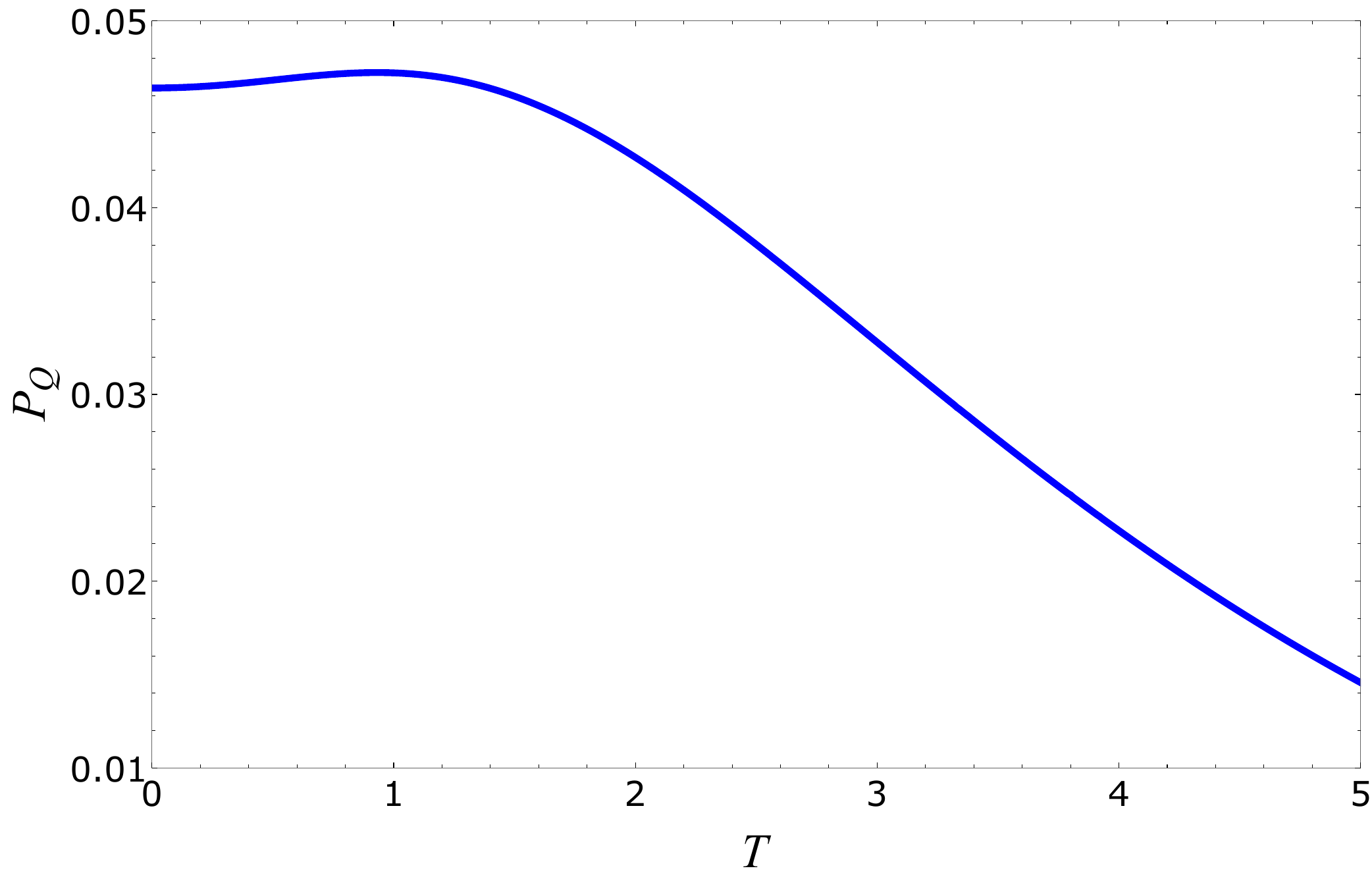}
\caption{Visibility for quantum noise ($P_Q$) vs. $T$. Parameters: $n = 4$, $ \hbar \omega_y = 1.365 meV$, $\hbar \omega_x =0.5 \hbar \omega_y$ and $\Delta T = 0.05 T$, where $T_C = 9K$.}
\label{figPQ}
\end{figure}

The conductance ratio in Fig.~\ref{fig10} and the quantum noise ratio shown in Fig.~\ref{fig11} exhibit very similar behavior, which limits their usefulness as probes of electron–hole symmetry breaking or mesoscopic quantum transport. Although both quantities display oscillations at energies corresponding to the QPC energy levels, their visibility,
$P_Q = \frac{\max\!\left(\frac{Q_{11}^{NQS}}{Q_{11}^{NQN}}\right)-\min\!\left(\frac{Q_{11}^{NQS}}{Q_{11}^{NQN}}\right)}{\max\!\left(\frac{Q_{11}^{NQS}}{Q_{11}^{NQN}}\right)+\min\!\left(\frac{Q_{11}^{NQS}}{Q_{11}^{NQN}}\right)}$,
remains relatively small and decreases further at higher energies (see, Fig. \ref{figPQ}), making it difficult to clearly resolve these transitions experimentally. This behavior arises because quantum noise contains both shot noise and {equilibrium noise} contributions, and in the temperature range considered here the thermal component dominates. Since {equilibrium noise} is closely related to conductance through fluctuation–dissipation relations, the ratio $\frac{Q_{11}^{\mathrm{NQS}}}{Q_{11}^{\mathrm{NQN}}}$ remains close to the value of $2$, similar to the conductance ratio, and therefore does not strongly reveal the effects of electron–hole symmetry breaking.

\endgroup

\begingroup

\subsection{$\Delta_T$ noise}
Figure~\ref{fig12}(a) presents a detailed comparison of the $\Delta_T$ noise in NQS and NQN junctions through the ratio 
$\Delta_T^{\mathrm{NQS}}/\Delta_T^{\mathrm{NQN}}$, plotted as a function of the dimensionless parameter 
$\frac{E_F - V_0}{\hbar \omega_x}$.
The corresponding numerical values are summarized in Table~\ref{Table3}. 
Our analysis shows that the upper bound of this ratio approaches 16 within the parameter regime considered here. 
At ultra–low temperature ($T = 0.01\,K$), the ratio approaches this theoretical upper bound. 
However, as the temperature increases, the ratio progressively decreases from this value.

The reduction of the ratio with temperature arises from the increasing importance of thermal excitations. 
At elevated temperatures, quasiparticle states above the superconducting gap become increasingly populated. 
As a result, the relative contribution of Andreev reflection to charge transport is reduced. 
Since the enhancement of $\Delta_T$ noise in the superconducting junction is strongly linked to Andreev processes, the suppression of these processes with increasing temperature leads to a corresponding reduction in the ratio 
$\Delta_T^{\mathrm{NQS}}/\Delta_T^{\mathrm{NQN}}$.

In Fig.~\ref{fig12}(a), we focus on the regime 
$1 \leq \frac{E_F - V_0}{\hbar \omega_x} \leq 10$, where the QPC introduces electron--hole asymmetry into the transport. 
Within this regime the {thermovoltage} and Seebeck coefficient become finite due to electron--hole symmetry breaking. 
Consequently, the ratio $\Delta_T^{\mathrm{NQS}}/\Delta_T^{\mathrm{NQN}}$ decreases with increasing temperature, as shown in Fig.~\ref{fig12}(a) and Table~\ref{Table3}.

To further clarify this behavior, we also examine a regime where electron--hole symmetry is effectively restored. 
Specifically, we consider the region 
$\frac{E_F - V_0}{\hbar \omega_x} \gg 10$, as shown in Fig.~\ref{fig12}(b). 
In this regime the QPC becomes nearly transparent and the transport is electron--hole symmetric. 
Under these conditions, the ratio 
$\Delta_T^{\mathrm{NQS}}/\Delta_T^{\mathrm{NQN}}$ reaches the exact value of 16 at $T = 0.01\,K$, consistent with the well–known theoretical upper bound. 
As the temperature increases, the ratio again decreases due to thermal effects. Here, we see that the visibility: $P_D=\frac{max\left(\frac{\Delta_T^{NQS}}{\Delta_T^{NQN}}\right)-min\left(\frac{\Delta_T^{NQS}}{\Delta_T^{NQN}}\right)}{max\left(\frac{\Delta_T^{NQS}}{\Delta_T^{NQN}}\right)+min\left(\frac{\Delta_T^{NQS}}{\Delta_T^{NQN}}\right)}$ is significantly larger unlike the conductance, quantum noise, see Fig. \ref{figPD}.

\begin{widetext}
 
\begin{figure}[H]
\centering
\includegraphics[scale=0.30]{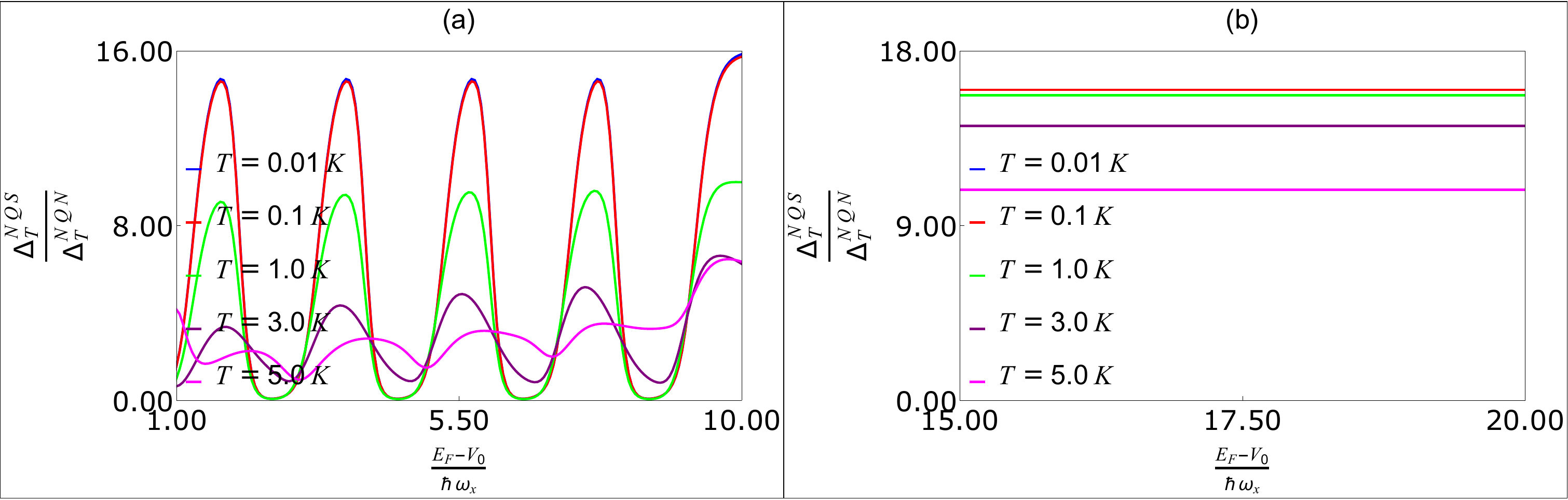}
\caption{Ratio of $\Delta_T$ noise $\frac{\Delta_{T}^{NQS}}{\Delta_{T}^{NQN}}$ for (a) e-h asymmetric, (b) e-h symmetric case vs. $\frac{E_F - V_0}{\hbar \omega_x}$. Parameters: $n = 4$, $ \hbar \omega_y = 1.365 meV$, $\hbar \omega_x =0.5 \hbar \omega_y$  and $\Delta T = 0.05T$, where $T_C = 9K$. {The curves corresponding to the temperatures $T = 0.01K,0.1K$ nearly overlap and are therefore difficult to distinguish on the scale of the figure}.}
\label{fig12}
\end{figure}
\end{widetext}

\begin{widetext}

\begin{table}[H]
\centering
\caption{Comparison of conductance, Seebeck coefficient, quantum noise, and $\Delta_T$ noise ratios for the NQS and NQN junctions at different temperatures. (-) denotes undefined.}
\resizebox{\textwidth}{!}{
\begin{tabular}{|c|c|c|c|c|c|c|c|c|c|c|}
\hline

\multirow{2}{*}{Ratio} 
& \multicolumn{5}{c|}{\begin{tabular}{c} e-h asymmetric \\ $1 \leq \frac{E_F-V_0}{\hbar \omega_x} \leq 10$ \end{tabular}}
& \multicolumn{5}{c|}{\begin{tabular}{c} e-h symmetric \\ $\frac{E_F-V_0}{\hbar \omega_x} \gg 10$ \end{tabular}} \\
\cline{2-11}

& $T=0.01K$ & $T=0.1K$ & $T=1K$ & $T=3K$ & $T=5K$
& $T=0.01K$ & $T=0.1K$ & $T=1K$ & $T=3K$ & $T=5K$ \\
\hline

$\frac{G_{NQS}}{G_{NQN}}$
& 2 & 1.998 & 1.97 & 1.68 & 1.40
& 2 & 2 & 2 & 2 & 2 \\
\hline

$\frac{S_{NQS}}{S_{NQN}}$
 & Zero & Zero & Zero & 0.04 & 0.25
& - & - & - & - & - \\
\hline

$\frac{Q_{11}^{NQS}}{Q_{11}^{NQN}}$
& 2 & 1.997 & 1.996 & 1.69 & 1.38
& 2 & 2 & 2 & 2 & 2 \\
\hline

$\frac{\Delta_T^{NQS}}{\Delta_T^{NQN}}$
& 14.72 & 14.69 & 9.53 & 4.85 & 3.15
& 16 & 16 & 15.71 & 14.14 & 10.84 \\
\hline

\end{tabular}}
\label{Table3}
\end{table}

\end{widetext}

This comparison provides a clear physical signature of how electron--hole asymmetry influences thermally induced noise in mesoscopic hybrid junctions.

\subsection{Why we focus on $\Delta_T$ noise?}

The above comparison reveals an important and rather striking advantage of analyzing $\Delta_T$ noise in superconducting hybrid junctions where e-h symmetry is broken. As summarized in Table~\ref{Table3}, the ratios of conductance and quantum noise between the NQS and NQN junctions remain close to the universal value of $2$ over a wide temperature range, whereas the ratio of $\Delta_T$ noise reaches a much larger value and approaches the upper bound of $16$ at very low temperatures. The origin of this qualitative difference lies in the fundamentally different physical information contained in these quantities.

\begin{figure}[H]
\centering
\includegraphics[scale=0.25]{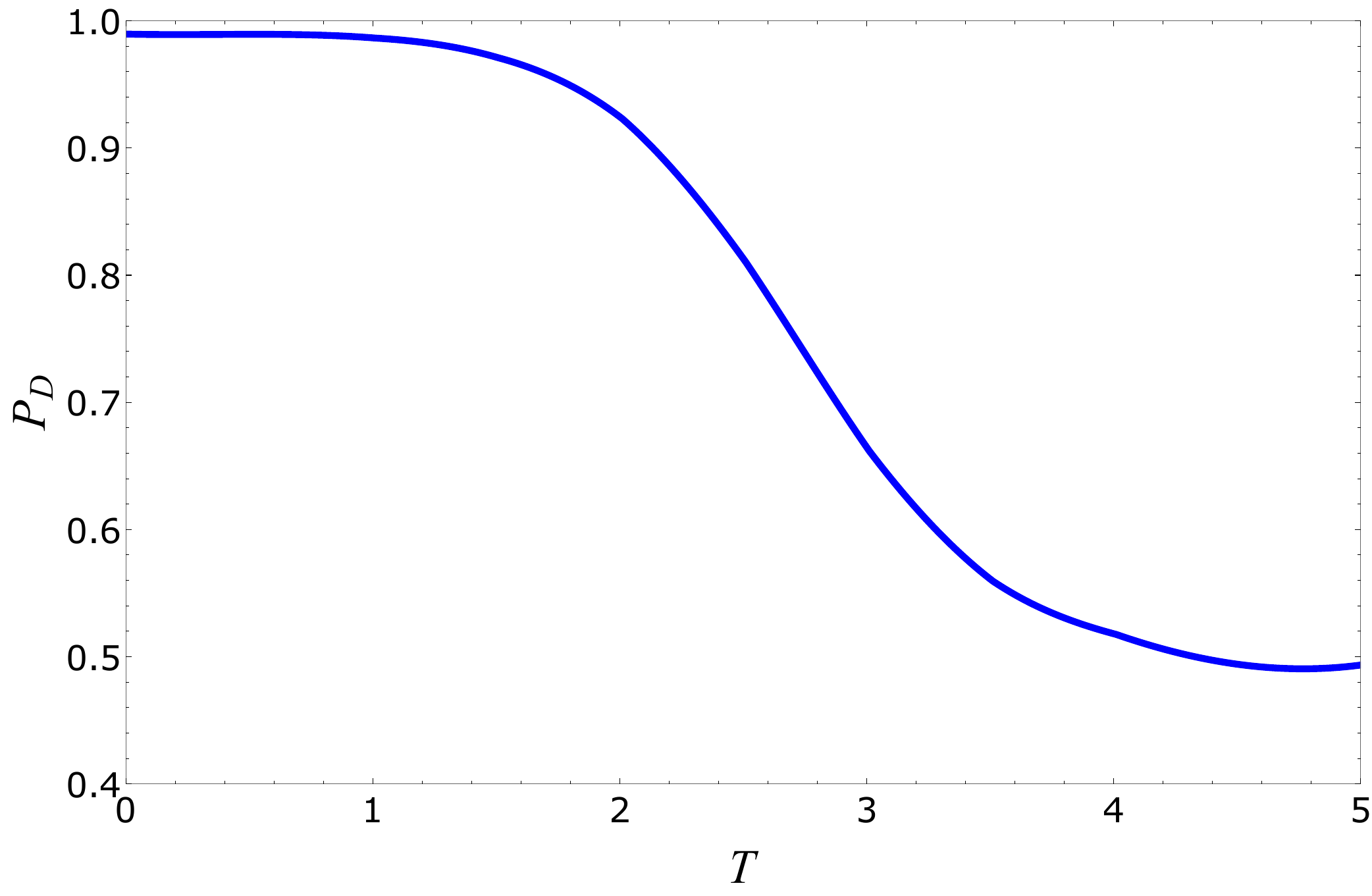}
\caption{Visibility for $\Delta_T$ noise ($P_D$) vs. $T$. Parameters: $n = 4$, $ \hbar \omega_y = 1.365 meV$, $\hbar \omega_x =0.5 \hbar \omega_y$ and $\Delta T = 0.05 T$, where $T_C = 9K$.}
\label{figPD}
\end{figure}

In contrast, $\Delta_T$ noise probes a fundamentally different physical regime. It measures current fluctuations generated purely by a temperature bias under the condition of zero average current. This constraint introduces a strong connection between $\Delta_T$ noise and thermoelectric transport processes. Thermoelectric effects arise only when electron--hole symmetry is broken giving rise to finite Seebeck coefficient and {thermovoltage}, because a perfectly e-h symmetric system results in zero thermoelectric response. When a quantum point contact breaks this symmetry, finite thermoelectric response is seen. In the superconducting junction, this asymmetry acts together with Andreev reflection processes, which intrinsically couple electron and hole excitations. The combined effect leads to a strong enhancement of $\Delta_T$ noise in the NQS junction relative to the NQN junction, producing the large ratio $\frac{\Delta_T^{\mathrm{NQS}}}{\Delta_T^{\mathrm{NQN}}}$ that approaches $16$ at very low temperatures.

An additional important observation emerges at intermediate temperatures such as $T = 1\,K$ and $T = 3\,K$. In this regime, the Seebeck coefficient and the corresponding {thermovoltage} become very small, see Fig. \ref{fig5}. Nevertheless, the $\Delta_T$ noise remains finite and continues to display a clear difference between the superconducting and normal junctions. This behavior highlights an important advantage of $\Delta_T$ noise as a probe of electron-hole symmetry breaking. Even when conventional thermoelectric quantities such as the Seebeck coefficient or {thermovoltage} become too small to provide a clear signal, the $\Delta_T$ noise remains sensitive to the underlying asymmetry in the transport channels. Therefore, $\Delta_T$ noise can reveal the presence and consequences of electron-hole symmetry breaking in regimes where standard thermoelectric measurements become ineffective.

Another important feature that emerges from Table~\ref{Table3} is the distinct temperature dependence of the different transport quantities. As the temperature increases, the ratios of conductance and quantum noise remain very close to the universal value of $2$, showing only a weak deviation from this value. This behavior reflects the fact that both conductance and quantum noise are primarily governed by Andreev reflection processes in the superconducting junction. Although thermal excitations gradually introduce quasiparticle contributions to transport, their effect remains relatively small and therefore the corresponding ratios do not change significantly with temperature. In contrast, the ratio $\frac{\Delta_T^{\mathrm{NQS}}}{\Delta_T^{\mathrm{NQN}}}$ shows a pronounced reduction from its upper bound of $16$ as temperature increases. This stronger temperature dependence arises because $\Delta_T$ noise is directly linked to thermoelectric transport driven by a temperature gradient. As thermal excitations become more prominent, the relative contribution of Andreev reflection is reduced and the resulting temperature-driven fluctuations decrease, leading to the observed suppression of the $\Delta_T$ noise ratio. This behavior clearly demonstrates that $\Delta_T$ noise is far more sensitive to electron--hole asymmetry than either conductance or quantum noise.

Another striking difference appears in the way the various quantities respond to the opening of successive QPC energy levels. As the parameter $\frac{E_F - V_0}{\hbar \omega_x}$ is varied, the system undergoes transitions between tunneling and transparent transport regimes corresponding to the opening of discrete QPC channels. For conductance, Seebeck coefficient, and quantum noise, these transitions produce only weak oscillations whose amplitudes remain relatively small. As a result, the changes associated with the opening of individual transport channels are not always clearly visible. In contrast, the ratio of $\Delta_T$ noise exhibits pronounced oscillatory behavior as each new channel becomes available for transport. These oscillations clearly track the quantum transport processes occurring across the different QPC energy levels, providing a much sharper signature of the underlying mesoscopic physics.

The origin of this enhanced sensitivity can be understood from the dependence of $\Delta_T$ noise on the transmission probability of the QPC. In mesoscopic transport theory, temperature-driven noise contains a factor proportional to $\mathcal{T}(1-\mathcal{T})$, where $\mathcal{T}$ denotes the net transmission probability of the junction. This functional form is maximized when the transmission is neither fully open nor fully closed, making $\Delta_T$ noise particularly sensitive to the transition between tunneling and transparent regimes. Consequently, as successive QPC energy levels open and the transmission changes rapidly, the $\Delta_T$ noise provides a clear and direct probe of these quantum transport processes in both NQN and NQS junctions.

Taken together, these results demonstrate that while conductance and conventional quantum noise primarily reflect the average charge transport and remain close to their universal superconducting enhancement factor of $2$, the $\Delta_T$ noise captures a richer set of physical effects arising from the interplay between Andreev reflection, thermoelectric transport, and electron--hole asymmetry. The strong enhancement, clear channel-resolved oscillations, and distinctive temperature dependence of the $\Delta_T$ noise therefore establish it as a powerful probe of symmetry breaking and mesoscopic quantum transport in superconducting quantum point contact systems.

\subsection{Reason behind considering $\hbar \omega_y$ of the same order as $\Delta_0$}

{The motivation for choosing $\hbar\omega_y$ to be of the same order as
$\Delta_0$ is to clearly resolve the interplay between the energy-dependent
electron-hole asymmetry introduced by the QPC and superconductivity. The
thermoelectric and noise properties investigated in the present work arise
from the competition between these two energy scales. If
$\hbar\omega_y\gg\Delta_0$, the transmission probability of the QPC varies only
weakly over the energy window relevant for superconducting transport, so that
the effect of electron-hole symmetry breaking on the transport properties
becomes significantly less pronounced. In contrast, when
$\hbar\omega_y\lesssim\Delta_0$, the transmission varies appreciably over the
same energy window, allowing the electron-hole symmetry breaking introduced by
the QPC to manifest itself clearly in the transport characteristics.
Consequently, choosing $\hbar\omega_y$ to be comparable to or less than $\Delta_0$ provides
the most appropriate regime for clearly identifying the interplay between
electron-hole symmetry breaking and superconductivity, which is the central
focus of the present work.}

{We have repeated the calculations using a substantially different
confinement energy, namely $\hbar\omega_y=\Delta_0/3$, where
$\Delta_0=1.76k_BT_C$ with $T_C=9\,K$, corresponding to
$\Delta_0=1.365\,\mathrm{meV}$, while keeping all other model parameters
unchanged. We find that the conductance exhibits quantized steps at
the same values of normalized Fermi energy as is obtained for the case $\hbar\omega_y=\Delta_0$. Consequently, the oscillatory features, and the underlying physical mechanisms remain
unchanged. The thermovoltage, quantum noise, and $\Delta_T$ noise also exhibit the same qualitative behavior and appear at the same transport-channel positions, with only quantitative changes in their magnitudes.}

{We have added the plots of conductance, thermovoltage, quantum noise and $\Delta_T$ noise, as well as the ratios of conductance, quantum noise and $\Delta_T$ noise for new parameter regime of $\hbar \omega_y = \Delta_0/3$, $\hbar \omega_x = \hbar \omega_y/2$, see Appendix \ref{AppD}.}

\endgroup
\section{Experimental realization and Conclusion}
\label{conclusion}

In this work, we have investigated the impact of electron–hole (e–h) symmetry breaking on the noise characteristics of mesoscopic hybrid junctions. Specifically, we focused on \( \Delta_T \) noise, in systems where a quantum point contact (QPC) serves as a tunable scatterer. By replacing the insulating barrier in conventional NIN and NIS junctions with a QPC, we deliberately break the intrinsic e–h symmetry, thereby enabling a deeper exploration of non-equilibrium transport fluctuations.

{The principal contribution of the present work is the establishment of a comprehensive theoretical framework for $\Delta_T$ noise in superconducting hybrid junctions with explicitly broken electron-hole symmetry. Unlike earlier studies of superconducting junctions, where electron-hole symmetry forces the thermovoltage to vanish and consequently restricts the evaluation of $\Delta_T$ noise at zero voltage, the energy-dependent transmission via the QPC generates a finite thermoelectric response, allowing the $\Delta_T$ noise to be evaluated self-consistently at finite thermovoltage. This previously unexplored regime reveals several new physical insights. First, the interplay between Andreev reflection and electron–hole symmetry breaking reduces the enhancement factor of the $\Delta_T$ noise in superconducting junctions relative to their normal counterparts compared with the enhancement observed in electron–hole symmetric superconducting junctions. While perfectly electron-hole symmetric superconducting junctions exhibit the universal enhancement factor of $16$, the explicit breaking of electron-hole symmetry suppresses this enhancement to approximately $14.70$ at $T=0.01$ and $0.1~K$, followed by a more pronounced reduction to $9.53$, $4.85$, and $3.15$ at $T=1$, $3$, and $5K$, respectively. Second, $\Delta_T$ noise exhibits pronounced channel-resolved oscillations associated with the opening of successive QPC transport modes, providing a much more sensitive probe of mesoscopic transport than conductance or conventional quantum shot noise. Finally, we demonstrate that even in temperature regimes where the Seebeck coefficient and thermovoltage become vanishingly small, $\Delta_T$ noise remains finite and continues to reveal the underlying electron-hole asymmetry. These results establish $\Delta_T$ noise as a powerful and complementary probe of nonequilibrium transport in superconducting quantum point contact junctions.}

{In previous studies, such as Refs. ~\cite{PhysRevLett.127.136801} and \cite{PhysRevB.107.075409}, consider only normal metallic contacts with a resonant-tunneling--type scatterer to
investigate the role of electron-hole asymmetry in $\Delta_T$ noise. In contrast, our work
focuses on the quantum point contact. Despite the different nature
of the scatterers, some qualitative similarities emerge in purely normal systems: the
{thermovoltage} remains negative throughout, and the resulting $\Delta_T$ noise is always
positive. However, while the analysis of $\Delta_T$ noise in junctions with only normal
metallic contacts is relatively straightforward, extending this study to superconducting
hybrid junctions constitutes a non-trivial problem due to the presence of many scattering processes.}

{In this manuscript, we consider a $s$-wave superconducting contact and demonstrate
that the interplay between Andreev reflection and electron--hole asymmetry leads to
qualitatively new physics. Most notably, we uncover that the ratio of $\Delta_T$ noise is always lower than the e-h symmetric cases. While $\Delta_T$ noise remains positive and exhibits
oscillations as a function of the normalized Fermi energy
$(E_F - V_0)/\hbar\omega_x$ in both normal and superconducting junctions, the magnitude of
the enhancement in the superconducting case is significantly larger due to Andreev
processes.}

Furthermore, we show that increasing temperature suppresses Andreev reflection and promotes transport via electron- and hole-like quasiparticle tunneling into the superconductor. As a result, the enhancement factor of $\Delta_T$ noise decreases from its maximal value. {Within the framework of the noninteracting Landauer-B\"uttiker scattering formalism considered in this article, the maximal enhancement factor of $16$ constitutes a universal bound for the class of hybrid junctions studied here. We emphasize that the inclusion of many-body interactions, charging effects, or other interaction-induced phenomena may modify this bound, which lies beyond the scope of the present work.} These results highlight the crucial role of superconductivity in shaping $\Delta_T$ noise and underscore the fundamentally different behavior of this noise in superconducting hybrid junctions compared to purely normal metallic contacts. We also show that the $\Delta_T$ noise provides a much more sensitive probe of the effects of e-h symmetry breaking on transport in a superconducting junction, which a Seebeck coefficient or the {thermovoltage} may not be able to provide.

{Within the framework of the noninteracting Landauer-Büttiker scattering formalism considered in this work, we find that the enhancement of the $\Delta_T$ noise in the NQS junction relative to the NQN junction is bounded by a maximum factor of 16. We emphasize that this bound is universal only for the class of noninteracting hybrid junctions considered here. The inclusion of many-body interactions, charging effects, or other interaction-induced phenomena may modify this bound, which lies beyond the scope of the present work.}

{It is important to note that in experiments the individual measurements of conductance, Seebeck coefficient, and noise may often contain several extrinsic contributions arising from device-specific parameters and measurement uncertainties \cite{PhysRevB.27.112, PhysRevLett.68.3765, jehl2000detection, kozhevnikov2000shot, PhysRevLett.84.3398}. These may include contact resistance, heating effects, calibration uncertainties in temperature gradients, and other setup-dependent effects. As a result, the absolute values of experimentally measured quantities can sometimes be influenced by such unwanted factors, making it difficult to directly identify the intrinsic physical mechanisms governing the transport. Taking ratios of the measured quantities between two closely related configurations, such as the NQS and NQN junctions considered here, provides a convenient way to suppress many of these extrinsic contributions. Since both measurements are performed within the same experimental setup and under nearly identical conditions, systematic errors and device-dependent prefactors largely cancel out when ratios are taken. Consequently, the resulting ratios highlight the intrinsic transport properties of the system more clearly and provide a more reliable probe of the underlying physics. In particular, the ratios considered in this work allow one to directly compare the influence of superconductivity and electron--hole symmetry breaking on the transport and noise characteristics of the junction, while minimizing the impact of experimental artifacts.}

The NQN junction we study here is inspired by earlier experimental and theoretical developments in QPC-based ballistic transport~\cite{BENENTI20171, PhysRevLett.60.848, wharam1988one, van1992thermo, PhysRevLett.68.3765}. For the superconducting segment, we consider a conventional $s$-wave superconductor with a critical temperature $T_C = 18$~K, corresponding to Nb \cite{PhysRev.95.1435}. While our setup considers a normal metal–QPC–superconductor (NQS) geometry, similar experimental architectures, such as superconductor–QPC–superconductor junctions, have been realized~\cite{bretheau2013exciting}. These can potentially be adapted to the NQS regime by replacing one of the superconducting terminals with a normal metal.

Finally, for experimental observation of $\Delta_T$ noise, one must isolate it from the {equilibrium noise} contribution at vanishing charge current. This can be achieved by performing two measurements: one at zero temperature bias (to obtain the thermal component at zero charge current), and another at finite temperature bias (to obtain zero current quantum noise). Subtracting the former from the latter yields the $\Delta_T$ noise~\cite{lumbroso2018electronic, shein2022electronic, sivre2019electronic}.

To the best of our knowledge, this is the first comprehensive theoretical study of $\Delta_T$ noise in mesoscopic normal-superconductor junctions that incorporates the effects of electron-hole symmetry breaking and studies the effect of finite {thermovoltage}. Our findings demonstrate that QPC-based hybrid structures offer a versatile and experimentally accessible platform for probing non-equilibrium fluctuation phenomena beyond the scope of conventional e-h symmetry preserving junctions.

\section*{Appendix}

The Appendix is organized into four parts. In Appendix~\ref{App_ch_Qn1}, we present the derivation of scattering amplitudes for both NQN and NQS junctions. Appendix~\ref{App_I} then presents the evaluation of the charge current flowing through the normal-metal lead for each of these junctions considered. Subsequently, we derive charge $\Delta_T$ {equilibrium noise} auto-correlation in Appendix \ref{App_ch_Qn}. 
% Finally, we add the Mathematica code to calculate charge $\Delta_T$ noise in SM \cite{suppl}.

\appendix
\label{appendix}

\begin{widetext}

\section{Scattering amplitudes in NQN and NQS junction}
\label{App_ch_Qn1}

The quantum point contact (QPC) was originally introduced in Ref. \cite{PhysRevB.41.7906} to study conductance quantization in the ballistic transport regime. Since then, it has been widely used in the study of thermoelectric effects~\cite{PhysRevLett.60.848, wharam1988one, van1992thermo, PhysRevLett.68.3765}. The potential landscape of the QPC is modeled by the following confinement potential~\cite{PhysRevB.41.7906}:
\begin{equation}
U(x, y) = \frac{1}{2}m^* \omega_y^2 y^2 - \frac{1}{2}m^* \omega_x^2 x^2 + V_0,
\end{equation}
where $m^*$ is electron's effective mass, and $V_0$ is the potential offset. The energy levels in the QPC are quantized as $E_n = V_0 + (n + 1/2)\hbar \omega_y$. The scattering amplitudes and the corresponding probabilities in a N-Q-N junction are already derived in Ref. \cite{PhysRevB.41.7906} and we follow exactly the same procedure and we do not repeat the derivation here. The corresponding transmission probability for the $n$th mode as derived in Ref. \cite{PhysRevB.41.7906} is given as
\begin{equation}\label{eqTQ}
\mathcal{T}_{Q} (E) =  \left(1 + e^{-2\pi(E + E_F - E_n)/\hbar \omega_x}\right)^{-1}.
\end{equation}

We use the expression of $\mathcal{T}_{Q} (E)$ further for the derivation of scattering amplitudes in NQS junction. We now derive the scattering amplitudes of a N-QPC-S junction, see Fig. \ref{fig15}. First, we consider a $N_1$-QPC-$N_2$-S junction, where $N_1$ stands for first normal metal and $N_2$ stands for second normal metal, after which we calculate $s$-matrix of N-QPC-S junction. For the derivation of scattering amplitudes, we closely follow \cite{PhysRevB.46.12841, RevModPhys.69.731, PhysRevB.99.134514}. First we consider the scattering matrices of N-QPC-N junction and N-S junction separately. Thus combining both the scattering matrices for $N_1$-QPC-$N_2$ and $N_2-S$ junction, we get the $s$-matrix for N-QPC-S junction in the limit, when length of the second normal metal tends to be zero. The incoming amplitudes $(c_{e}^{+}(N_1), c_e^-(N_2), c_h^-(N_1), c_h^+(N_2))$ and the outgoing amplitudes $(c_e^-(N_1), c_e^+(N_2), c_h^+(N_1), c_h^-(N_2))$ for $N_1$-QPC-$N_2$ junction are related by a scattering matrix, which is given as

\begin{figure}[H]
\centering
\includegraphics[scale=0.4]{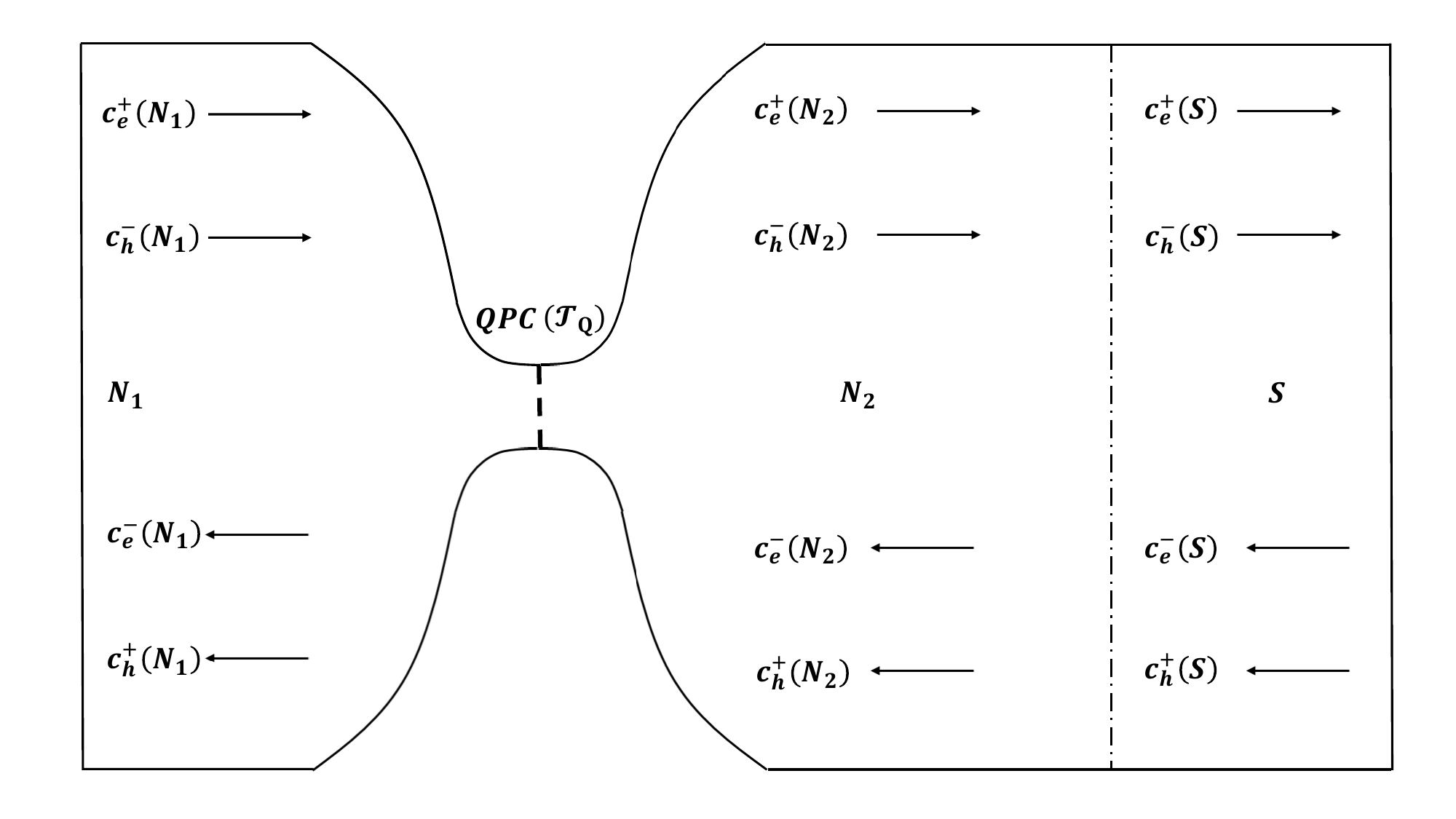}
\caption{Schematic diagram of the $N_1-Q-N_2-S$ junction. The black dashed lines between $N_2-S$ junction represent the interface between the normal metal (N) and the superconductor (S), while that in constriction between $N_1$ and $N_2$ represent the QPC with transmission probability $\mathcal{T}_Q$ (Eq. (\ref{eqTQ}). {Here, the $N_1-Q-N_2-S$ junction is considered first in order to calculate the scattering amplitudes in the NQS junction as shown in Fig. \ref{fig1}(b) by later putting the length of the second normal metal $N_2$ to be zero.}}
\label{fig15}
\end{figure} 

\begin{equation}
    S_{N_1 - QPC - N_2} = \begin{pmatrix}
        -i \sqrt{1 - \mathcal{T}_{Q}(E)} & \sqrt{\mathcal{T}_{Q}(E)} & 0 & 0\\
        \sqrt{\mathcal{T}_{Q}(E)} & -i \sqrt{1 - \mathcal{T}_{Q}(E)} & 0 & 0\\
      0 & 0 &  -i \sqrt{1 - \mathcal{T}_{Q}(-E)} & \sqrt{\mathcal{T}_{Q}(-E)}\\
      0 & 0 &  \sqrt{\mathcal{T}_{Q}(-E)} & -i \sqrt{1 - \mathcal{T}_{Q}(-E)}
    \end{pmatrix}
\end{equation}

Similarly, at the perfectly transparent $N_2-S$ junction, the incoming amplitudes $(c_e^+(N_2), c_e^-(S), c_h^-(N_2), c_h^+(S))$ and the outgoing amplitudes $(c_e^-(N_2), c_e^+(S), c_h^+(N_2), c_h^-(S))$ are related by the scattering matrix, which is given as \cite{PhysRevB.25.4515, mishra2025delta_t},

\begin{equation}
    S_{N_2-S} = \begin{pmatrix}
         0 & \sqrt{|u_0|^2 - |v_0|^2} c & a & 0\\
         \sqrt{|u_0|^2 - |v_0|^2}c & 0 & 0 & -a\\
         a & 0 & 0 & \sqrt{|u_0|^2 - |v_0|^2}c^*\\
         0 & -a & \sqrt{|u_0|^2 - |v_0|^2}c^* & 0
    \end{pmatrix}
\end{equation}

where $a = \frac{v_0}{u_0}$, $c = \frac{1}{u_0}$, with $u_0(v_0) = \left[\frac{1}{2} \{1 \pm \frac{\sqrt{E^2 - \Delta^2}}{E}\}\right]^{\frac{1}{2}}$ for $E > \Delta$ and for $E < \Delta$, we get $u_0(v_0) = \left[\frac{1}{2} \{1 \pm \frac{i \sqrt{\Delta^2 - E^2}}{E}\}\right]^{\frac{1}{2}}$. One can connect the incoming amplitudes $(c_e^+(N_1), c_e^-(S), c_{h}^-(N_1), c_h^+(S))$ and outgoing amplitudes $(c_e^-(N_1), c_e^+(S), c_{h}^+(N_1), c_h^-(S))$ of $N_1$ and $S$ and one can get the scattering amplitudes of the N-Q-S junction. The scattering matrix of the NQS junction is given as

\begin{equation}
    S_{NQS} = \begin{pmatrix}
        s_{11}^{ee}(E) & s_{12}^{ee}(E) & s_{11}^{eh}(E) & s_{12}^{eh}(E) \\
        s_{21}^{ee}(E) & s_{22}^{ee}(E) & s_{21}^{eh}(E) & s_{22}^{eh}(E) \\
        s_{11}^{he}(E) & s_{12}^{he}(E) & s_{11}^{hh}(E) & s_{12}^{hh}(E) \\
        s_{21}^{he}(E) & s_{22}^{he}(E) & s_{21}^{hh}(E) & s_{22}^{hh}(E)
    \end{pmatrix},
\end{equation}

where $s_{l k}^{\rho \gamma}(E)$ is the amplitude for a particle of type $\gamma \in \{e, h \}$ to scatter from terminal $k \in \{1,2 \}$ to terminal $l \in \{1,2 \}$ as a particle of type $\rho \in \{e, h \}$ at energy $E$. The elements of the scattering matrix are given as

\begin{equation}
    \begin{split}
        s_{11}^{ee}(E) &= \frac{i \left(\sqrt{1-\mathcal{T}_{Q}(E)}-a^2 \sqrt{1-\mathcal{T}_{Q}(-E)}\right)}{D}, \quad s_{12}^{ee}(E) = -\frac{c \sqrt{\mathcal{T}_{Q}(E)}}{D}, \quad s_{11}^{eh}(E) = -\frac{a \sqrt{\mathcal{T}_{Q}(E)} \sqrt{\mathcal{T}_{Q}(-E)}}{D}, \\
         s_{12}^{eh}(E) &= -\frac{i a c^* \sqrt{\mathcal{T}_{Q}(E)} \sqrt{1-\mathcal{T}_{Q}(-E)}}{D}, \quad s_{21}^{ee}(E) = -\frac{c \sqrt{\mathcal{T}_{Q}(E)}}{D}, \quad s_{21}^{eh}(E) = \frac{i a c \sqrt{1-\mathcal{T}_{Q}(E)} \sqrt{\mathcal{T}_{Q}(-E)}}{D},  \\
         s_{22}^{ee} (E)&= \frac{i c^2 \sqrt{1-\mathcal{T}_{Q}(E)}}{D}, \quad s_{22}^{eh}(E) = \frac{a \left(1-c c^* \sqrt{1-\mathcal{T}_{Q}(E)} \sqrt{1-\mathcal{T}_{Q}(-E)}\right)-a^3 \sqrt{1-\mathcal{T}_{Q}(E)} \sqrt{1-\mathcal{T}_{Q}(-E)}}{D},\\
         s_{11}^{he}(E) &= -\frac{a \sqrt{\mathcal{T}_{Q}(E)} \sqrt{\mathcal{T}_{Q}(-E)}}{D}, \quad s_{12}^{he}(E) = \frac{i a c \sqrt{1-\mathcal{T}_{Q}(E)} \sqrt{\mathcal{T}_{Q}(-E)}}{D},\quad s_{21}^{hh}(E) = -\frac{c^* \sqrt{\mathcal{T}_{Q}(-E)}}{D},  \\
         s_{12}^{hh}(E) &= -\frac{c^* \sqrt{\mathcal{T}_{Q}(-E)}}{D},\quad s_{21}^{he}(E) = -\frac{i a c^* \sqrt{\mathcal{T}_{Q}(E)} \sqrt{1-\mathcal{T}_{Q}(-E)}}{D}, \quad
        s_{11}^{hh}(E) = \frac{i \left(a^2 \sqrt{1-\mathcal{T}_{Q}(E)}-\sqrt{1-\mathcal{T}_{Q}(-E)}\right)}{D},  \\ s_{22}^{hh}(E) &= -\frac{i c^{* 2} \sqrt{1-\mathcal{T}_{Q}(-E)}}{D},\quad s_{22}^{he}(E) = -\frac{a^3 \sqrt{1-\mathcal{T}_{Q}(E)} \sqrt{1-\mathcal{T}_{Q}(-E)}+a \left(c c^* \sqrt{1-\mathcal{T}_{Q}(E)} \sqrt{1-\mathcal{T}_{Q}(-E)}-1\right)}{D}.
    \end{split}
\end{equation}

where $D = a^2 \sqrt{1 - \mathcal{T}_{Q}(E)} \sqrt{1 - \mathcal{T}_{Q}(-E)} - 1$. We will use the scattering matrix elements to further derive the charge current as well as a general expression of quantum noise, which comprises quantum {equilibrium noise} and quantum shot noise. Later, we focus on the quantum shot noise evaluated at a finite thermovoltage, defined at zero net charge current in the normal metal, i.e., charge $\Delta_T$ noise.

\section{Average charge current in NQN and NQS junction}
\label{App_I}

In mesoscopic conductors, the average current entering terminal $\alpha$ can be expressed within the scattering matrix framework as $\langle I_{\alpha} \rangle = \frac{2e}{h} \sum_{\beta} \int_{-\infty}^{\infty} dE , (\delta_{\alpha\beta} - \mathcal{T}_{\alpha\beta}) f_{\beta}$. Here, $\mathcal{T}_{\alpha\beta}$ denotes the probability that an electron incident from lead $\beta$ is transmitted into lead $\alpha$, with the summation over $\beta$ running over all connected terminals. The numerical factor of 2 accounts for spin degeneracy. The occupation of electronic states in terminal $\beta$ is described by the Fermi–Dirac function $f_{\beta} = \left(1 + e^{\frac{E - e V_{\beta}}{k_B T_{\beta}}}\right)^{-1}$, where $V_{\beta}$ represents the voltage applied to terminal $\beta$ and the corresponding temperature is given by $T_{\beta} = T + \Delta T_{\beta}$, with $T$ denoting the reference equilibrium temperature and $\Delta T_{\beta}$ specifying the temperature bias at that terminal $\beta$.
For the specific case of an NQN junction involving two normal leads labeled by $\beta \in \{1,2\}$, the expression for the current flowing into terminal 1 simplifies to $\langle I_1 \rangle = \frac{2e}{h} \int_{-\infty}^{\infty} dE \, \mathcal{T}_{Q}(E) \bigl(f_{1e} - f_{2e}\bigr)$. Here, $\mathcal{T}_{Q}(E)$ represents the energy-dependent transmission probability across the junction. 

Similarly, in a multiterminal superconducting hybrid junction consisting of normal metal and superconductor, the charge current out of a normal metallic terminal is given as \cite{BENENTI20171},

\begin{equation}
    \begin{split}
        \langle I_{\alpha} \rangle &= \frac{2 e }{h} \sum_{\beta} \sum_{\rho, \xi \in \{e, h \}} \int_0^{\infty} dE \,\,\, sgn(\rho) \left[\delta_{\alpha \beta} \delta_{\rho \xi} - \mathcal{T}_{\alpha \beta}^{\rho \xi}\right] f_{\beta}^{\xi}(E).
    \end{split}
\end{equation}

For each terminal labeled by $\beta \in \{1,2\}$, the occupation probability of quasiparticle states of type $\xi \in {e,h}$ is described by the distribution function $f_{\beta}^{\xi}(E)=\left[1+\exp\left(\frac{E+\mathrm{sgn}(\xi)V_{\beta}}{k_B T_{\beta}}\right)\right]^{-1}$, where the sign function assigns $\mathrm{sgn}(\xi)=+1$ to electron-like excitations and $\mathrm{sgn}(\xi)=-1$ to hole-like excitations, with the same convention applied to the indices $\rho$ and $\xi$. Making use of these definitions, the expression for the charge current flowing through the normal lead of an NQS junction, as introduced in Ref.~\cite{BENENTI20171}, can be recast into a simplified form:

\begin{equation}
\langle I_1\rangle = \frac{2 e }{h} \sum_{\beta} \sum_{\rho, \xi \in \{e, h \}} \int_0^{\infty} dE \,\,\, sgn(\rho) \left[\delta_{1 \beta} \delta_{\rho \xi} - \mathcal{T}_{1 \beta}^{\rho \xi}\right] f_{\beta}^{\xi}(E).
\end{equation}
Now, the charge current in NQS junction is given as,

\begin{equation}
\langle I_1^{NQS} \rangle = \frac{2e}{h} \int_0^{\infty} dE (1 + \mathcal{R}^A_e (E) - \mathcal{R}^B_e (E)) (f_{1e} - f_{2e}) + \frac{2e}{h} \int_0^{\infty} dE (1 + \mathcal{R}^A_h (E) - \mathcal{R}^B_h (E)) (f_{1h} - f_{2e}) 
\end{equation}

where $\mathcal{R}^A_e (E) = |s_{11}^{he}(E)|^2$ and $\mathcal{R}^B_e (E) = |s_{11}^{ee}(E)|^2$, $\mathcal{R}^A_h (E) = |s_{11}^{eh}(E)|^2$ and $\mathcal{R}^B_h (E) = |s_{11}^{hh}(E)|^2$. Now, utilizing the property, i.e., $s_{11}^{eh}(E) = s_{11}^{he *}(-E)$, $s_{11}^{hh}(E) = s_{11}^{ee *}(-E)$, and $f_{1h}(-E) = 1 - f_{1e}(E)$, we get the total charge current to be,

\begin{equation}
    \langle I_1^{NQS} \rangle = \frac{2e}{h} \int_{-\infty}
^{\infty} dE \, (1 + \mathcal{R}^A(E) - \mathcal{R}^B (E)) (f_{1e} - f_{2e}),
\end{equation}

where $\mathcal{R}^A (E) = \mathcal{R}^A_e (E)$, $\mathcal{R}^B (E) = \mathcal{R}^B_e (E)$.

\section{Charge quantum noise and $\Delta_T$ noise in NQN and NQS junctions}
\label{App_ch_Qn}
The general expression of charge quantum noise $Q_{ij}$ is given as \cite{BENENTI20171, martin2005course}

\begin{eqnarray}
Q_{ij} &=& \frac{2e^2}{h} \int \sum_{p,q \in \{1,2\}} A_{p;l}(i,E) A_{q;k}(j,E) \left[f_p(E)(1 - f_q(E)) + f_q(E)(1 - f_p(E))\right] dE 
\end{eqnarray}

Here, $A_{p; l}(i, E) = \delta_{ip} \delta_{il} - s_{ip}^{\dagger} s_{il}$. The terms $Q_{11\, \mathrm{eq}}^{NQN}$ and $Q_{11\, \mathrm{sh}}^{NQN}$ correspond to quantum {equilibrium} and quantum shot noise contributions, respectively.

Total charge quantum noise in NQN junction is $Q^{NQN}_{11}=Q^{NQN}_{11sh}+Q^{NQN}_{11\mathrm{eq}}$, where the charge shot noise contribution ($Q^{NQN}_{11sh}$) and charge {equilibrium noise} contribution ($Q^{NQN}_{11\mathrm{eq}}$) are given as,
\begin{equation}
\begin{split}
Q^{NQN}_{11sh} &=\frac{4e^2}{h} \int^{\infty}_{-\infty} dE \left[ \left\{ \mathcal{T}_{Q}(E)(1-\mathcal{T}_{Q}(E)) \right\} \left(f_{1e} - f_{2e} \right)^2 \right] = \frac{4e^2}{h} \int^{\infty}_{-\infty} F^{NQN}_{11sh} \left(f_{1e}- f_{2e} \right)^2 dE, \\
\text{and}~~Q^{NQN}_{11\mathrm{eq}} &= \frac{4e^2}{h} \int^{\infty}_{-\infty} dE \left[ \mathcal{T}_{Q}(E)\left(\sum_{i \in \{1,2 \}} f_{ie}(1-f_{ie}) \right) \right] =\frac{4e^2}{h} \int^{\infty}_{-\infty} F^{NQN}_{11\mathrm{eq}} \left(\sum_{i \in \{1,2 \}} f_{ie}(1-f_{ie}) \right) dE,
\label{B_NN}
\end{split}
\end{equation}
where the scattering terms depend on the QPC transmission probability $\mathcal{T}_{Q}(E)$, and $F^{NQN}_{11sh}=\mathcal{T}_{Q}(E)~( 1 - \mathcal{T}_{Q}(E)) $ while $F^{NQN}_{11\mathrm{eq}}= \mathcal{T}_{Q}$. 

$\Delta_T$ noise refers to the quantum shot noise-like component at finite temperature bias and vanishing current. For NQN junction, $\langle I_1^{NQN} \rangle$ is zero at finite {thermovoltage} $V_{th}^{NQN}$. Once we evaluate $V_{th}^{NQN}$, one can calculate $Q_{11sh}^{NQN}$, which is nothing but $\Delta_T$ noise, which is written as $\Delta_T^{NQN}$ for the NQN junction. Therefore, the expression for $\Delta_T^{NQN}$ is exactly same as $Q_{11sh}^{NQN}$ as written in Eq. (\ref{B_NN}) and is given as

\begin{eqnarray}
    \Delta_T^{NQN} &=\frac{4e^2}{h} \int^{\infty}_{-\infty} dE \left[ \left\{ \mathcal{T}_{Q}(E)(1-\mathcal{T}_{Q}(E)) \right\} \left(f_{1e} - f_{2e} \right)^2 \right] = \frac{4e^2}{h} \int^{\infty}_{-\infty} F^{NQN}_{11sh} \left(f_{1e}- f_{2e} \right)^2 dE
\end{eqnarray}

The general expression of zero frequency charge quantum noise $Q_{i j}(\omega=0$) \cite{PhysRevB.53.16390}, in a hybrid superconducting junction is written as,
\begin{eqnarray}
Q_{ij} && = \frac{2e^2}{h} \int \sum_{ \substack{p,q \in \{1, 2\} ,\\
\mu,\nu,\sigma,\rho \in \{e,h\}} } \mathrm{sgn}(\mu) sgn(\nu)  
A_{p,\sigma;q,\rho}(i \mu) 
A_{q,\rho;p,\sigma}(j \nu) 
({f}_{p \sigma} [1-{f}_{q \rho}]+(1-{f}_{p \sigma}) {f}_{q \rho}) \, dE,
\label{eqn:sn}
\end{eqnarray}
where $A_{p,\sigma;q,\rho}(i \mu) 
= \delta_{i p} \delta_{i q} \delta_{\mu \sigma} \delta_{\mu \rho} 
- s^{\mu \sigma *}_{i p} s^{\mu \rho}_{i q}$. 
Here, for electron $\mathrm{sgn}(\mu)=sgn(\nu)=+1$ and for hole, they are -1. Charge noise auto-correlation (i.e., $i=j=1$), i.e., charge  quantum noise in the normal metal of a NQS junction, is,
\begin{eqnarray}
Q^{NQS}_{11} && = \frac{2e^2}{h} \int \sum_{ \substack{p,q \in \{1, 2\} ,\\
\mu,\nu,\sigma,\rho \in \{e,h\}} } \mathrm{sgn}(\mu) sgn(\nu)  
A_{p,\sigma;q,\rho}(1 \mu) 
A_{q,\rho;p,\sigma}(1 \nu) 
({f}_{p \sigma} [1-{f}_{q \rho}]+(1-{f}_{p \sigma}) {f}_{q \rho}) \, dE,
\end{eqnarray}

Using the $s$-matrix as shown in Eq. (B3), charge quantum noise auto-correlation can be simplified as,
\begin{eqnarray}
Q^{NQS}_{11} &=& Q_{11\mathrm{eq}}^{NQS} + Q_{11sh}^{NQS},
\label{B_Snoise}
\end{eqnarray}
where,

\begin{equation}
    \begin{split}
        Q_{11\mathrm{eq}}^{NQS} &= \frac{4e^2}{h} \int_{-\infty}^{\infty} dE \bigg[(1 - \mathcal{R}^B(E) + \mathcal{R}^A(E))^2 + \mathcal{R}^A(-E) \mathcal{R}^B(E) + \mathcal{R}^A(E) \mathcal{R}^B(-E) + \mathcal{R}^B(E)(\mathcal{T}^C(E) + \mathcal{T}^D(-E)) \\& + \mathcal{R}^A(E)(\mathcal{T}^D(E) + \mathcal{T}^C(-E) + 2 \mathcal{R}^A(E) \mathcal{R}^B(E))\bigg]   f_{1e}(1-f_{1e}) + \frac{4e^2}{h} \int_{-\infty}^{\infty} dE \bigg[\mathcal{T}^C(E)^2 + \mathcal{T}^D(E)^2 \\& + 2 \mathcal{T}^C(E) \mathcal{T}^D(-E) + \mathcal{R}^B(E) (\mathcal{T}^C(E) + \mathcal{T}^D(-E)) + \mathcal{R}^A(E)(\mathcal{T}^C(-E) + \mathcal{T}^D(E))\bigg]f_{2e}(1-f_{2e}),\\
        Q_{11sh}^{NQS} &= \frac{4e^2}{h} \int_{-\infty}^{\infty}dE \bigg[\mathcal{R}^B(E) (\mathcal{T}^C(E) + \mathcal{T}^D(-E)) + \mathcal{R}^A(E) (\mathcal{T}^C(-E) + \mathcal{T}^D(E)) + 2 \mathcal{R}^A(E) \mathcal{R}^B(E) \\&+ 2 Re(s_{11}^{ee}(E) s_{11}^{ee * }(-E) s_{11}^{he * }(E) s_{11}^{he  }(-E)) \bigg]  (f_{1e} - f_{2e})^2 + \frac{4e^2}{h} \int_{-\infty}^{\infty}dE \bigg[\mathcal{R}^A(E) \mathcal{R}^B(E)\\& -  Re(s_{11}^{ee}(E) s_{11}^{ee * }(-E) s_{11}^{he * }(E) s_{11}^{he  }(-E)) \bigg](f_{1e} - f_{1h})^2.
    \end{split}
    \label{B_NS}
\end{equation}

{Here, $Q_{11\mathrm{eq}}^{NQS}$ and $Q_{11sh}^{NQS}$ are the {equilibrium} and shot noise-like component in NQS junction. The various scattering mechanisms are described in terms of their associated probabilities, where Andreev reflection is represented by $\mathcal{R}^A(E) = |s_{11}^{he}(E)|^2$ and the normal reflection probability is given by $\mathcal{R}^B(E) = |s_{11}^{ee}(E)|^2$. In addition, the transmission of quasiparticles across the junction is quantified by $\mathcal{T}^C(E) = |s_{21}^{ee}(E)|^2$ for electron-like excitations and by $\mathcal{T}^D(E) = |s_{21}^{he}(E)|^2$ for hole-like excitations, both evaluated at energy $E$. Similarly, one can also calculate the scattering probabilities such as $\mathcal{R}^A(-E)$, $\mathcal{R}^B(-E)$, $\mathcal{T}^C(-E)$ and $\mathcal{T}^D(-E)$ by replacing $E$ by $-E$ in the expressions of scattering probabilities.}

Here, $\Delta_T$ noise for NQS junction is same as $Q_{11sh}^{NQS}$ at non-zero temperature bias and vanishing current. In the NQS junction, the charge current $\langle I_1^{NQS} \rangle = 0$ is achieved at {thermovoltage} $V_{th}^{NQS}$. Then, $Q_{11sh}^{NQS}$ can be evaluated at finite $V_{th}^{NQS}$, which is equal to $\Delta_T^{NQS}$. Therefore, the expression for $\Delta_T^{NQS}$ is exactly same as $Q_{11sh}^{NQS}$ as written in Eq. (\ref{B_NS}) and is given as

\begin{equation}
\begin{split}
    \Delta_T^{NQS} &=\frac{4e^2}{h} \int_{-\infty}^{\infty}dE \bigg[\mathcal{R}^B(E) (\mathcal{T}^C(E) + \mathcal{T}^D(-E)) + \mathcal{R}^A(E) (\mathcal{T}^C(-E) + \mathcal{T}^D(E)) + 2 \mathcal{R}^A(E) \mathcal{R}^B(E) \\&+ 2 Re(s_{11}^{ee}(E) s_{11}^{ee * }(-E) s_{11}^{he * }(E) s_{11}^{he  }(-E)) \bigg]  (f_{1e} - f_{2e})^2 + \frac{4e^2}{h} \int_{-\infty}^{\infty}dE \bigg[\mathcal{R}^A(E) \mathcal{R}^B(E)\\& -  Re(s_{11}^{ee}(E) s_{11}^{ee * }(-E) s_{11}^{he * }(E) s_{11}^{he  }(-E)) \bigg](f_{1e} - f_{1h})^2.
        \end{split}
\end{equation}

\section{Conductance, Seebeck Coefficient, Quantum noise and $\Delta_T$ noise at $\hbar \omega_y = \Delta_0/3$}
\label{AppD}
\begingroup

To demonstrate that the results presented in the main text are not an artifact
of the particular choice of the QPC confinement energy, we repeat all the
calculations for a substantially different value of the transverse confinement
energy, namely $\hbar\omega_y=\Delta_0/3$, where
$\Delta_0=1.76k_BT_C$ with $T_C=9~K$, corresponding to
$\Delta_0=1.365~\mathrm{meV}$. All other model parameters are kept identical to
those used throughout the main text.

The corresponding conductance, Seebeck coefficient, quantum
noise, $\Delta_T$ noise, together with the ratios
$G_{\mathrm{NQS}}/G_{\mathrm{NQN}}$,
$Q_{11}^{\mathrm{NQS}}/Q_{11}^{\mathrm{NQN}}$, and
$\Delta_T^{\mathrm{NQS}}/\Delta_T^{\mathrm{NQN}}$, are shown in
Figs.~\ref{fig22}--\ref{fig28}. It is evident that changing the confinement
energy by a factor of three does not alter any of the qualitative conclusions
presented in the main text. In particular, the conductance continues to exhibit
the same quantized step structure, while the Seebeck coefficient, quantum noise, and $\Delta_T$ noise display the same
overall behavior as functions of the normalized Fermi energy. Furthermore, the
oscillatory features associated with the successive opening of QPC transport
channels occur at the same positions as those presented in the main text,
demonstrating that the underlying transport mechanisms remain unchanged.

A comparison of the corresponding ratios further confirms the robustness of our
results. The ratios of conductance and quantum noise continue to approach
their maximal value of $2$, while the ratio of $\Delta_T$ noise reaches the
same upper bound of $16$ obtained in the main text. Thus, the characteristic
enhancement factors arising from superconductivity remain completely
unchanged. Only small quantitative differences are observed, primarily at
higher temperatures where thermally excited quasiparticles contribute more
significantly to transport. These changes affect only the numerical values of
the transport coefficients and noise, without modifying their qualitative
behavior or the underlying physical interpretation.

These results clearly demonstrate that the conclusions of the present work are
robust against substantial variations in the transverse confinement energy of
the QPC. The phenomena discussed throughout the manuscript originate from the
interplay between the energy-dependent transmission of the QPC and Andreev
reflection in the superconducting contact, rather than from any particular
choice of $\hbar\omega_y$. Consequently, the principal findings of this work,
including the oscillatory behavior of the transport quantities, the temperature
dependence of the superconducting enhancement, and the universal enhancement
factors of $2$ for conductance and quantum noise and $16$ for
$\Delta_T$ noise, remain invariant under a significant change of the QPC
confinement energy.

\endgroup

\begin{figure}[H]
\centering
\includegraphics[scale=0.30]{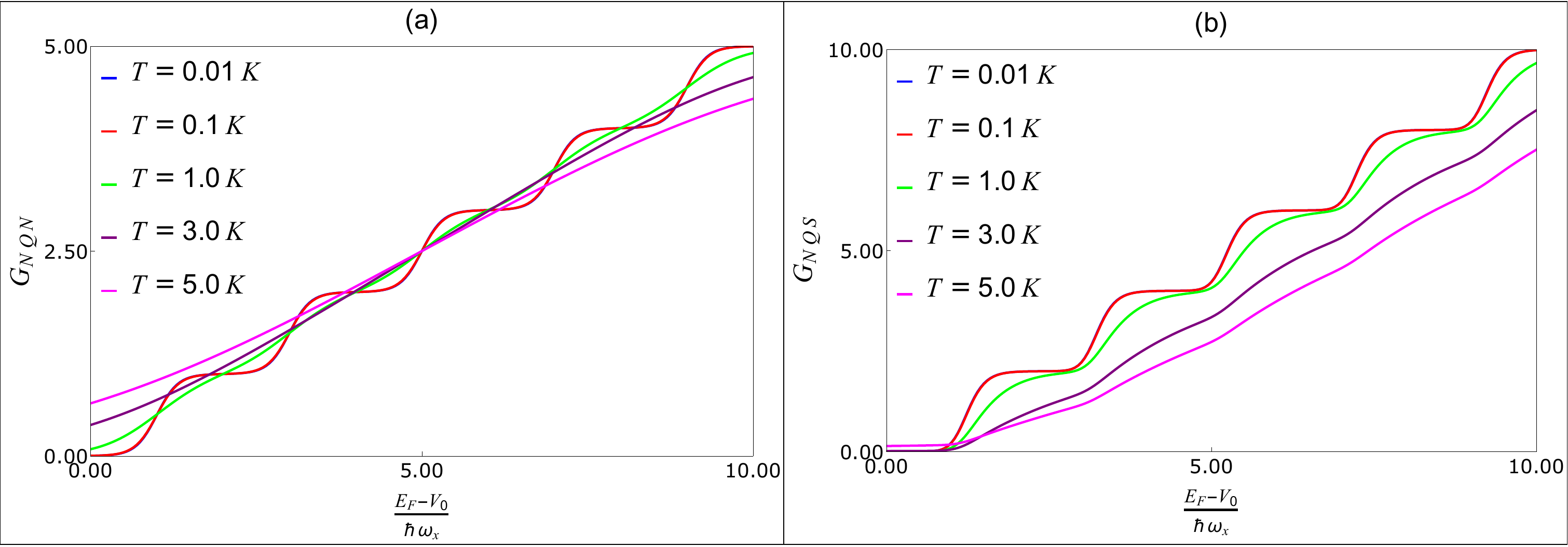}
\caption{Charge conductance in (a) NQN ($G_{NQN}$) and (b) NQS ($G_{NQS}$) (in units of $\frac{2e^2}{h}$) vs. normalized Fermi energy ($\frac{E_F - V_0}{\hbar \omega_x}$) at different temperatures considered above. Parameters are $ \hbar \omega_y = 0.455 meV$, $\hbar \omega_x =0.5 \hbar \omega_y$, $n=4$, with $T_C = 9K$. {Curves corresponding to the lowest temperatures nearly overlap and are therefore difficult to distinguish on the scale of the figure}.}
\label{fig22}
\end{figure}

\begin{figure}[H]
\centering
\includegraphics[scale=0.30]{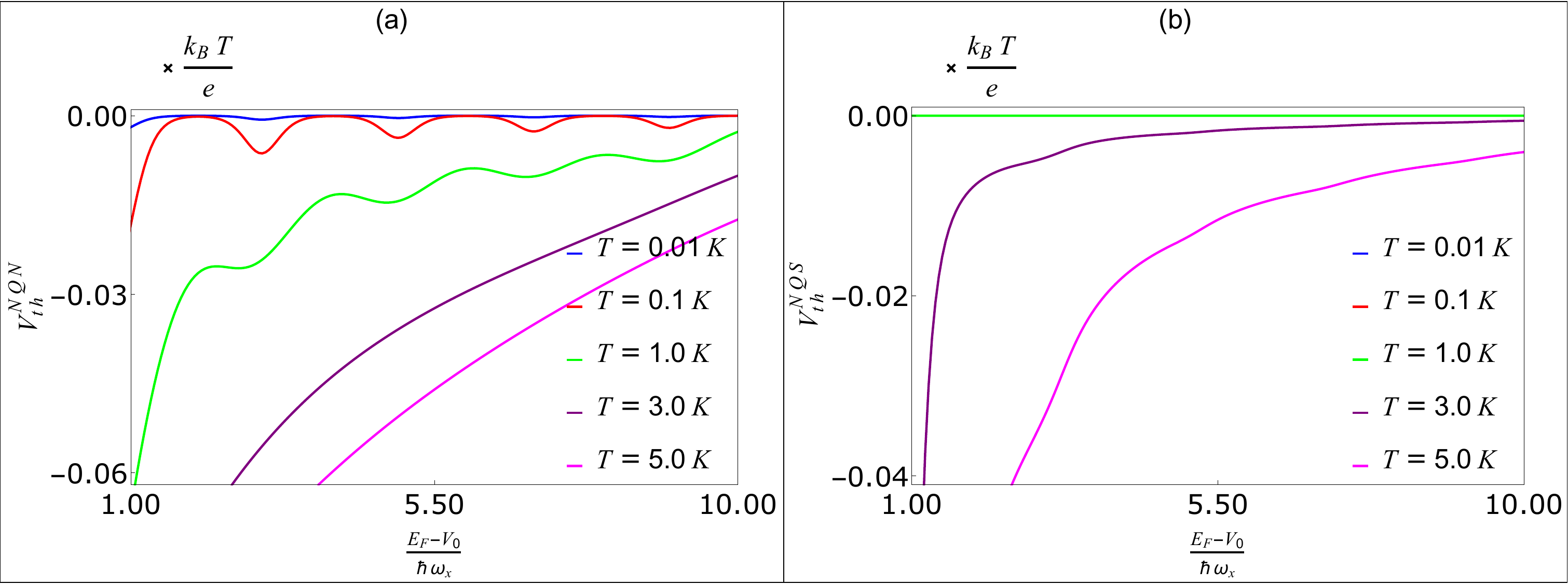}
\caption{{Thermovoltage} in (a) NQN ($V^{NQN}_{th}$) and (b) NQS ($V_{th}^{NQS}$) (in units of $\frac{k_B T}{e}$) vs. normalized Fermi energy $\frac{E_F - V_0}{\hbar \omega_x}$. Parameters are $n = 4$, $ \hbar \omega_y = 0.455 meV$, $\hbar \omega_x =0.5 \hbar \omega_y$ and $\Delta T = 0.05 T$, where $T_C = 9K$. {Curves corresponding to the lowest temperatures in (b) and (d) nearly overlap and are therefore difficult to distinguish on the scale of the figure}.}
\label{fig23}
\end{figure}

\begin{figure}[H]
\centering
\includegraphics[scale=0.30]{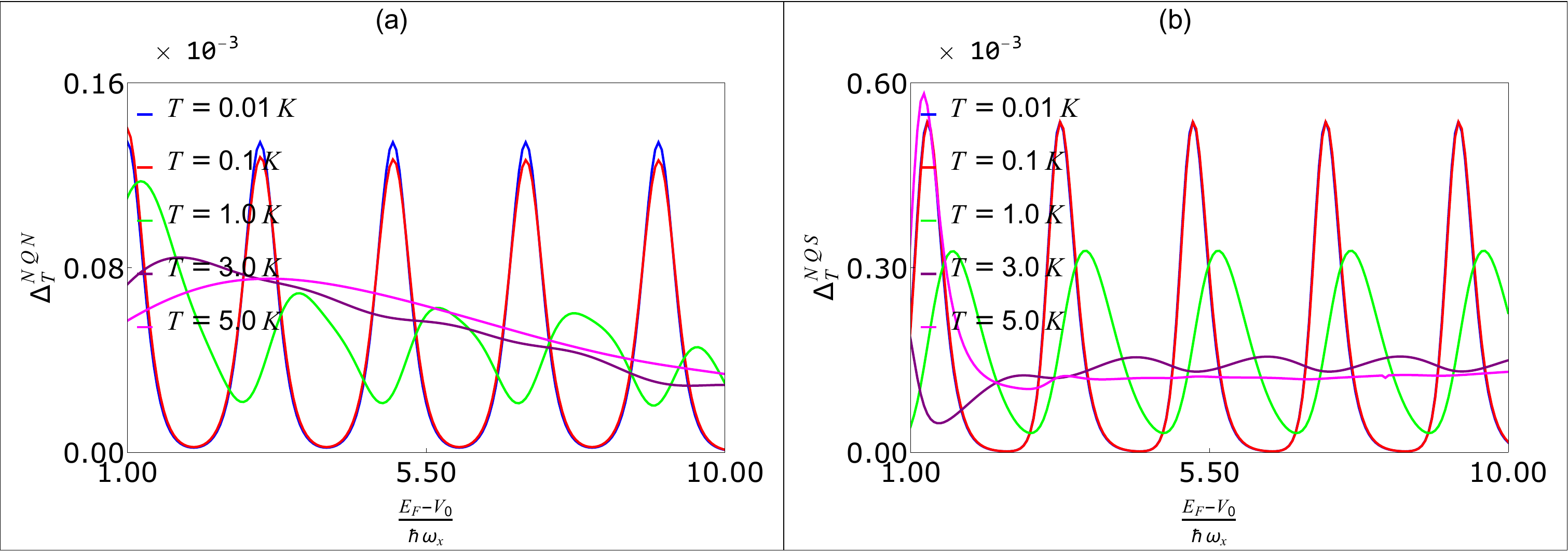}
\caption{Charge $\Delta_T$ noise in (a) NQN ($\Delta^{NQN}_{T}$) and (b) NQS ($\Delta_{T}^{NQS}$) (in units of $\frac{4e^2}{h} k_B T$) vs. normalized Fermi energy $\frac{E_F - V_0}{\hbar \omega_x}$. Parameters taken are $n = 4$, $ \hbar \omega_y = 0.455 meV$, $\hbar \omega_x =0.5 \hbar \omega_y$ and $\Delta T = 0.05 T$, where $T_C = 9K$. {The curves corresponding to the lowest temperatures nearly overlap and are therefore difficult to distinguish on the scale of the figure}.}
\label{fig24}
\end{figure}

\begin{figure}[H]
\centering
\includegraphics[scale=0.30]{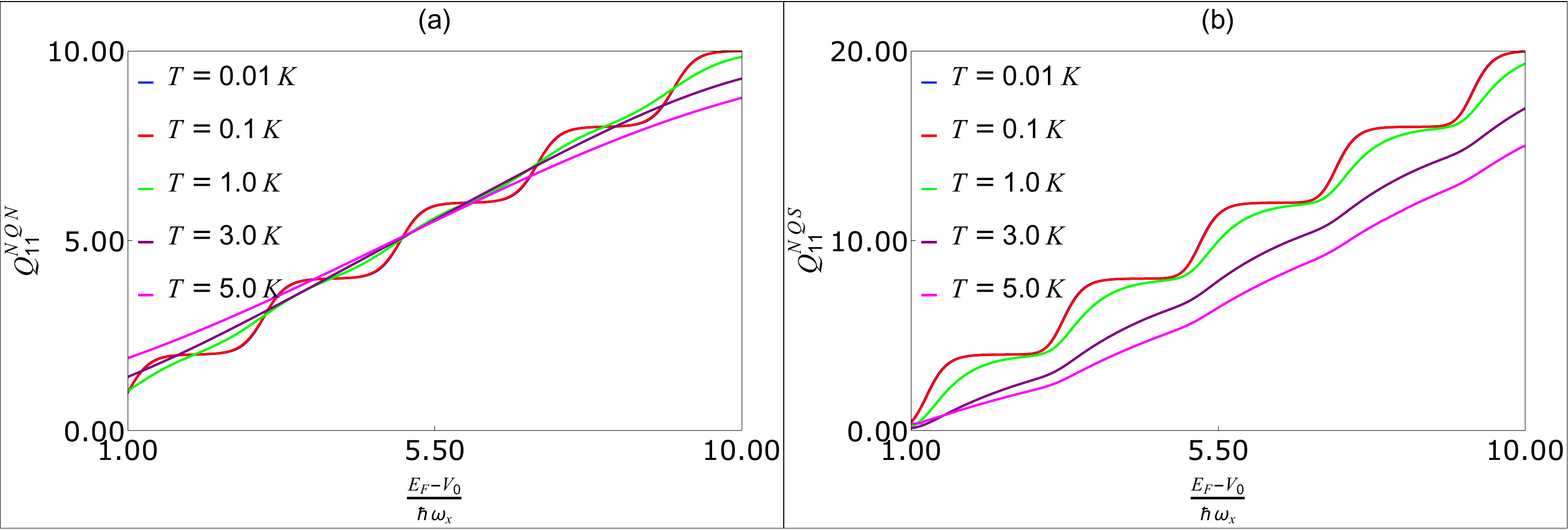}
\caption{Quantum noise in (a) NQN ($Q^{NQN}_{11}$) and (b) NQS ($Q_{11}^{NQS}$) (in units of $\frac{4e^2}{h} k_B T$) vs. normalized Fermi energy $\frac{E_F - V_0}{\hbar \omega_x}$. Parameters are $eV = 0.1k_BT$, $n = 4$, $ \hbar \omega_y = 0.455 meV$, $\hbar \omega_x =0.5 \hbar \omega_y$ and $eV=0.1k_B T$, where $T_C = 9K$. {The curves corresponding to the lowest temperatures nearly overlap and are therefore difficult to distinguish on the scale of the figure}.}
\label{fig25}
\end{figure}

\begin{figure}[H]
\centering
\includegraphics[scale=0.30]{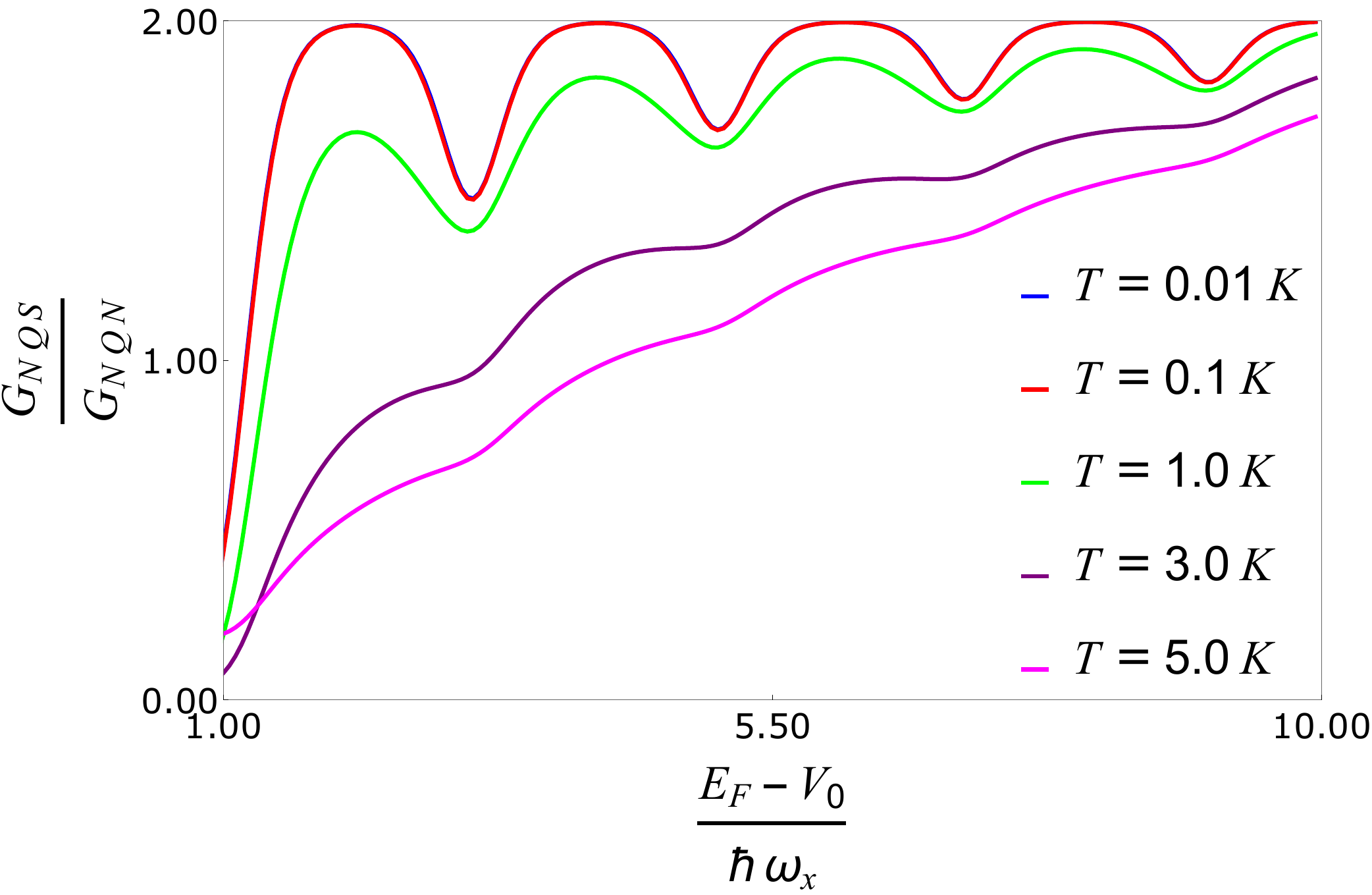}
\caption{Ratio of conductance $\frac{G_{NQS}}{G_{NQN}}$ for (a) e-h asymmetric, (b) e-h symmetric case vs. $\frac{E_F - V_0}{\hbar \omega_x}$. Parameters are $n = 4$, $ \hbar \omega_y = 0.455 meV$, $\hbar \omega_x =0.5 \hbar \omega_y$ and $\Delta T = 0.05 T$, where $T_C = 9K$.  {The curves corresponding to the lowest temperatures nearly overlap in (a) and are therefore difficult to distinguish on the scale of the figure. All curves in (b) nearly overlap and are therefore difficult to distinguish}.}
\label{fig26}
\end{figure}

\begin{figure}[H]
\centering
\includegraphics[scale=0.30]{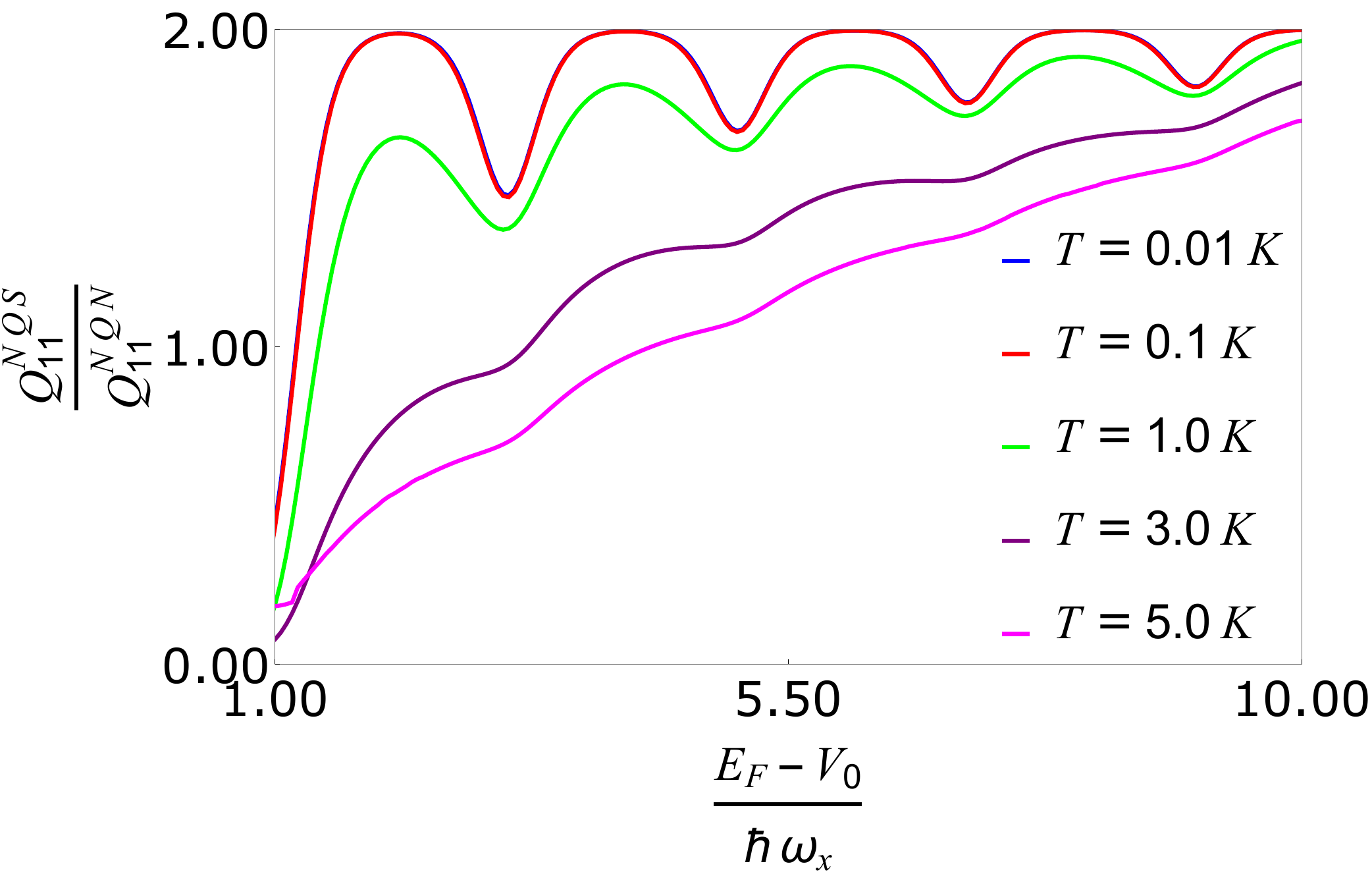}
\caption{Ratio of quantum noise $\frac{Q_{11}^{NQS}}{Q_{11}^{NQN}}$ for vs. $\frac{E_F - V_0}{\hbar \omega_x}$. Parameters: $n = 4$, $ \hbar \omega_y = 0.455 meV$, $\hbar \omega_x =0.5 \hbar \omega_y$ and $eV=0.1k_B T$, where $T_C = 9K$. {The curves corresponding to the lowest temperatures nearly overlap in (a) and are therefore difficult to distinguish on the scale of the figure. All curves in (b) nearly overlap and are therefore difficult to distinguish.}}
\label{fig27}
\end{figure}

\begin{figure}[H]
\centering
\includegraphics[scale=0.30]{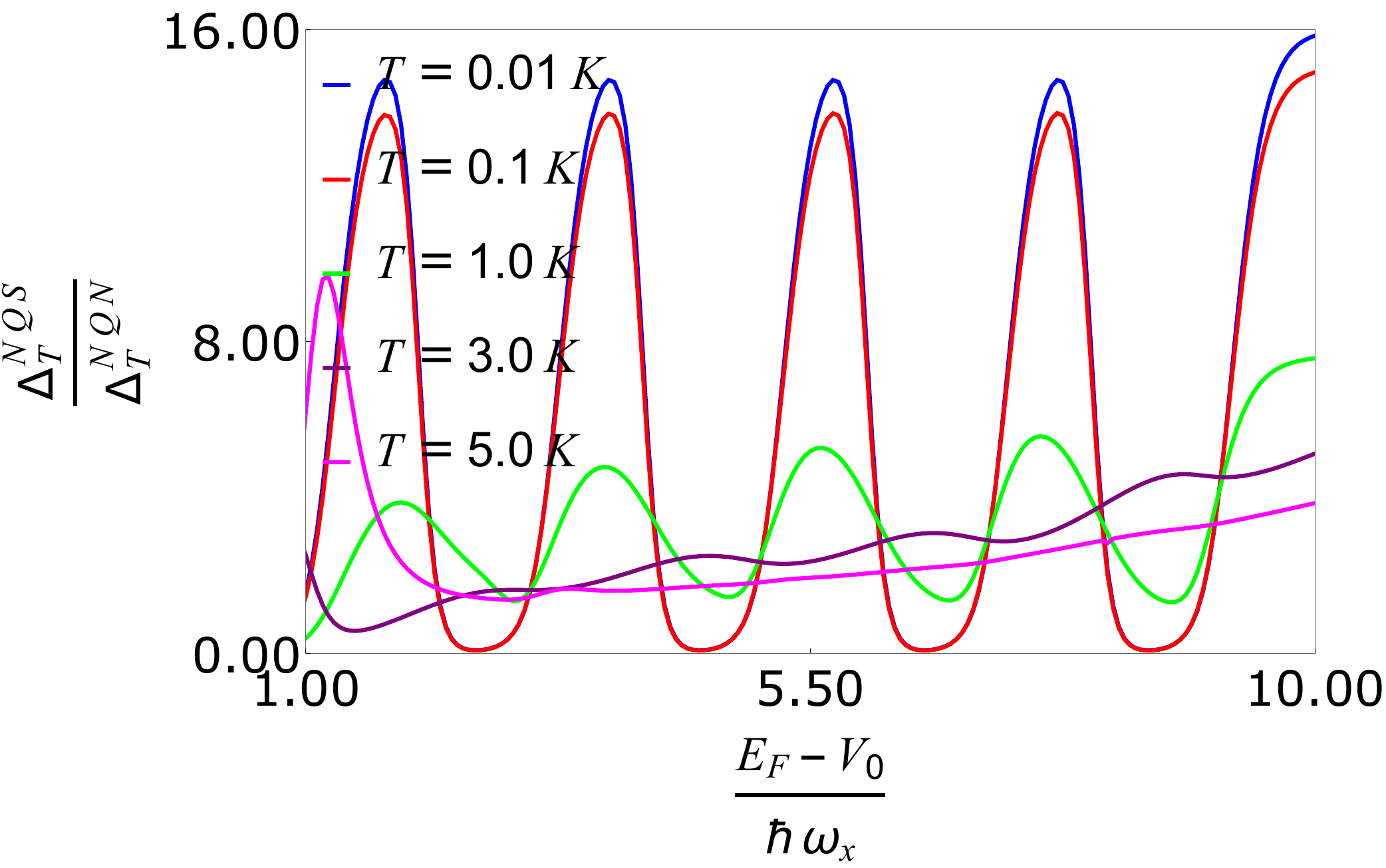}
\caption{Ratio of $\Delta_T$ noise $\frac{\Delta_{T}^{NQS}}{\Delta_{T}^{NQN}}$ vs. $\frac{E_F - V_0}{\hbar \omega_x}$. Parameters: $n = 4$, $ \hbar \omega_y = 0.455 meV$, $\hbar \omega_x =0.5 \hbar \omega_y$  and $\Delta T = 0.05T$, where $T_C = 9K$. {The curves corresponding to the lowest temperatures nearly overlap and are therefore difficult to distinguish on the scale of the figure}.}
\label{fig28}
\end{figure}

\end{widetext}

\bibliography{apssamp5}

\end{document}